\documentclass[12pt]{article}
\usepackage{graphicx}

\begin{document}
\def\be{\begin{equation}}
\def\ee{\end{equation}}
\def\bfi{\begin{figure}}
\def\efi{\end{figure}}
\def\bea{\begin{eqnarray}}
\def\eea{\end{eqnarray}}
\newcommand{\ket}[1]{\vert#1\rangle}
\newcommand{\bra}[1]{\langle#1\vert}
\newcommand{\braket}[2]{\langle #1 \vert #2 \rangle}
\newcommand{\ketbra}[2]{\vert #1 \rangle \! \langle #2 \vert}
\def\id{\mathbb{I}}
\title{Aging in Domain Growth}
\author{Marco Zannetti \\
Dipartimento di Fisica ``E.R.Caianiello''\\
Universit\`a di Salerno, Italy}
\date{}
\maketitle

\section{Introduction}

Slow relaxation and aging are ubiquitous in the low temperature behavior of
complex systems, notably structural and spin glasses~\cite{Bouchaud,CugliandoloLH}. 
The challenge of modern out of
equilibrium statistical mechanics is to make the theory of these phenomena. 
In such a context, domain growth is of particular importance as a paradigmatic example where
the mechanisms leading to slow relaxation and aging are believed to be well understood.
In this chapter aging in domain growth will be reviewed, showing, however, that even in
this case the off-equilibrium behavior is far from trivial and that more work is needed
to reach a full understanding.

\begin{figure}[h]
\begin{center}
\includegraphics*[scale = 0.5]{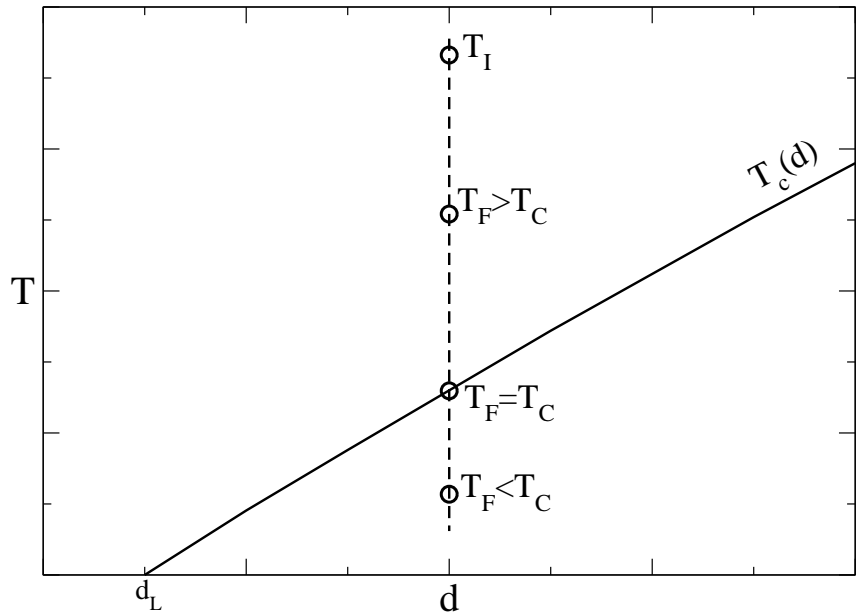}
\end{center}
\caption{\label{fig1}Generic phase diagram of a phase-oredering system on the $(d,T)$ plane. 
$d_L$ is the lower critical dimensionality.}
\end{figure}

As it has been explained in the first chapter of this book,
domain growth takes place after the quench to below the critical point 
of a phase-ordering system, such as a ferromagnet or a binary mixture. However, in order
to have a comprehensive view of the problem, it is useful to let
the final temperature of the quench to span the
whole temperature range, from above to below the critical temperature. More precisely,
consider the typical phase diagram in the dimensionality-temperature plane, as schematically
drawn in Fig.\ref{fig1}. The critical temperature depends on the space dimensionality $d$ of the
system. Although the specific form of the function $T_C(d)$ varies from model to
model, generic features are that there exists a lower critical 
dimensionality $d_L$, such that $T_C(d)=0$ for $d \leq d_L$ and that there is
a monotonous increase of $T_C(d)$ for $d > d_L$.
The value of $d_L$ depends on the symmetry of the model, with $d_L=1$ for
discrete symmetry and $d_L=2$ for continous
symmetry. The disordered and ordered phases are above and below the critical line $T_C(d)$,
respectively.

Let us first see where the slow relaxation comes from.
The system is supposed to be initially in equilibrium at some temperature
$T_I$ well above the critical line. For simplicity, take $T_I=\infty$ corresponding 
to an initial state with vanishing correlation length. For a certain
value of the dimensionality $d>d_L$ (dashed vertical line in Fig.\ref{fig1}), the system may be quenched 
to a final value of the temperature $T_F$ 
greater, equal or lower than the critical temperature $T_C$. The relaxation
involves the growth of the time dependent correlation length $\xi(t)$ from the initial value $\xi_I=0$ to a
final value $\xi_F>0$, which depends on $T_F$. Specifically, $0<\xi_F < \infty$ for $T_F>T_C$,
while $\xi_F=\infty$ for $T_F \leq T_C$ if the system is infinite 
(it will be explained below why $\xi_F=\infty$ also when
$T_F < T_C$). Therefore, for $T_F>T_C$ there is a finite equilibration time $t_{eq}$. 
For $T_F \leq T_C$, since $\xi(t)$ typically grows with a power law $\xi(t) \sim t^{1/z}$~\cite{Bray94}, the system is 
always out of equilibrium no matter for how long it is
observed. Hence, for phase-ordering systems the mechanism responsible of the slow relaxation 
in the quench to or to below $T_C$ is clear, it is the
growth of correlated regions of arbitrarily large size.

The next step is to focus on the features of the relaxation, like aging,
which depend on where the system is quenched to on the $(d,T)$ plane. 
Assuming that the istantaneous
quench takes place at $t=0$, aging is manifested through the behavior of
two times observables, such as the order parameter autocorrelation function
$C(t,t_w)$ and the autoresponse function $R(t,t_w)$. The shortest time after the 
quench $t_w \geq 0$ and the longest $t \geq t_w$ are conventionally called the waiting time
and the observation time. Regarding $t_w$ as the age of the system, aging is usually, and
quite effectively, meant to say that older systems relax slower and younger ones faster~\cite{Struick}. However,
this needs to be explained in more detail, since a rich phenomenology
goes under the same heading of aging. 

The first relevant feature
is the separation of the time scales~\cite{CugliandoloLH}. This means
that when $t_w$ is sufficiently large, the range of 
$\tau = t-t_w$ is devided into the short $\tau \ll t_w$
and the long $\tau \gg t_w$ time separations, and that quite different
behaviors are observed in the two regimes. For short $\tau$
the system appears equilibrated. The two time quantities
are time translation invariant (TTI) and exhibit the same behavior as if equilibrium at 
the final temperature of the quench had been reached
\be
C(t,t_w) = C_{eq}(\tau,T_F), \;\;\;\; R(t,t_w) = R_{eq}(\tau,T_F).
\label{I.1}
\ee
This is the quasi-equilibrium regime, where the system is ageless.
Conversely, for large $\tau$ there is a
genuine off-equilibrium behavior obeying a scaling form called simple aging~\cite{Bouchaud,CugliandoloLH}
\be
C(t,t_w) = C_{ag}(t,t_w) = t_w^{-b}h_C(t/t_w)
\label{I.2}
\ee
\be
R(t,t_w) = R_{ag}(t,t_w) = t_w^{-(1+a)}h_R(t/t_w)
\label{I.3}
\ee
where $a,b$ are non negative exponents and $h_{C,R}(t/t_w)$ are scaling functions. 
Here, the additional important features are i) that the system does not realize, so to speak, 
that it is off-equilibrium untill $\tau \sim t_w$ and ii) that once the off-equilibrium 
relaxation gets started, the time scale of the relaxation is fixed
by $t_w$ itself. Hence, the system ages.

Aging is common to all quenches: above, 
to and to below $T_C$. In the first case, where the equilibration time is finite,
for aging to be observable it is necessary that $t_{eq}$ is sufficiently large to 
allow for both a large $t_w$ and $t_w \ll t_{eq}$, so that the separation of the time scales
is possible. Eventually, when $t_w$ hits $t_{eq}$, equilibrium is reached and aging is interrupted.

The phenomenology outlined above
suggests the existence of {\it fast} and {\it slow} degrees of freedom~\cite{Mazenko88,Franz,EPJ}. 
The separation of the time
scales, then, means that, for sufficiently large $t_w$, the fast degrees of
freedom have already thermalized, while the slow degrees of freedom are still out of
equilibrium. This picture is easy to visualize in the case of domain growth. Thinking,
for simplicity, of a ferromagnetic system, the fast degrees of freedom are responsible of the thermal
fluctuations within ordered domains, while the slow ones are the labels of domains,
like the spontaneous magnetization within each domain. The label fluctuates since a given site at
different times may belong to different domains. For a given $t_w$, the typical
size of a domain is $L(t_w) \sim t_w^{1/z}$ and it takes an interval of time
$\tau \sim t_w$ for a domain wall to sweep the whole domain. Hence, in the short time regime
the slow degrees of freedom are frozen and only the fast degrees of freedom 
contribute to the decay of the two time quantities, yielding the behavior~(\ref{I.1}).
Conversely, in the long time regime, the time evolution is dominated by the motion of
domain walls, producing the off-equilibrium behavior of Eqs.~(\ref{I.2}) and~(\ref{I.3}). 

The picture just outlined is simple and intuitive enough to justify 
the opinion that aging in domain growth is well understood. Indeed, in
section \ref{largeN} an example will be presented where the construction
of the fast and slow degrees of freedom can be carried out exactly.
However, if the picture works well for the quenches to below $T_C$, 
it cannot be so easily extended also to the quenches to $T_C$, where 
the interpretation in terms of fast and slow degrees
of freedom remains less clear. Despite the common features in the phenomenology,
a qualitative difference between these two instances of aging,
whose origin has not yet been satisfactorily clarified, arises in the way the matching between
the stationary and the aging behavior is implemented. In the critical quench stationary and aging behaviors
match {\it multiplicatively}, while in the quench to below $T_C$ the matching is {\it additive}. 

The existence 
of these two different realizations of aging poses, among others, quite an interesting problem
for what happens in the quench to the special final state located at $(d=d_L, T_F=0)$.
Looking at Fig.\ref{fig1}, it is evident that such a state can be regarded as a limit state
reached either along the critical line or from the ordered region, as $d \rightarrow d_L$.
However, the two points of view are not equivalent, because the structure of the two times
quantities in one case ought to be multiplicative, while in the other additive. As we shall
see, the second alternative is the correct one. Nonetheless, in the quench to $(d=d_L, T_F=0)$
there are peculiar features which make it a case apart from both the quenches to and to below
$T_C$ with $d > d_L$. As a matter of fact, the processes listed above can be hierarchically
organized in terms of increasing degree of deviation from equilibrium. At the bottom there is
the quench to $T_F>T_C$, where $t_{eq}$ is finite. Immediately above there is the quench to
$T_C$, where it takes an infinite time to reach equilibrium. Still above there
is the quench to $T_F < T_C$ with $d>d_L$, where equilibrium is not reached even in
an infinite time. However, this is revealed by the autocorrelation function without appearing
in the dynamic susceptibility, which instead behaves as if equilibrium was reached.
Lastly, at the top of the hierarchy there is the quench to $(d=d_L, T_F=0)$, where also the dynamic
susceptibility displays out of equilibrium behavior over all time scales.

For what concerns the computation of the aging properties, namely the exponents $a,b$ and the scaling
functions $h_{C,R}(t,t_w)$, the systematic expansion methods of field theory, like the $\epsilon$-expansion,
can be succesfully used in the quench to $T_C$~\cite{Janssen}\cite{Calabrese}. Perturbative
methods, instead, are useless in the quench to below $T_C$, where the best theoretical tool
for analytical calculations remains the uncontrolled Gaussian auxiliary field (GAF) approximation
of the Ohta-Jasnow-Kawasaki type~\cite{Ohta}, in its various formulations~\cite{mazenko89}. 
With methods of this type
a good understanding of the autocorrelation function $C(t,t_w)$ has been achieved~\cite{Liu}
(see also the first chapter),
while the computation of the autoresponse function $R(t,t_w)$, in particular of the exponent $a$,
remains very much an open problem. For this reason, the investigation of aging in the quench to below $T_C$
relies heavily on numerical simultions, although the accurate numerical computation of $R(t,t_w)$
is a difficult problem of its own~\cite{Chat1,noialg,corrsc}. In order to complete
the theoretical panorama, the local scale invariance hypothesis~\cite{LSI} must be mentioned. 
This is a conjecture according to which the response function transforms covariantly under
the group of local scale transformations, both in the quenches to and to below $T_C$.
However, the predictions of this theory are affected by discrepancies with the renormalization
group calculations at $T_C$~\cite{Calabrese} and with the numerical simulations at 
and below $T_C$, which seem to indicate~\cite{Pleimling}\cite{Crquench} that the local scale 
invariance hypothesis is akin
to an approximation of Gaussian nature. Nonetheless, a definite assessment of this  
approach cannot yet be made, since work is in progress with the proposal of modified
versions of the theory~\cite{newLSI}.

In the following sections an overview of the aging properties in the various quenches mentioned above will
be presented, first in general and then through analytical and numerical results for specific models.
Preliminary to this discussion is a short summary of the static properties.

\section{Statics}

In general, order parameter configurations will be denoted by $[\varphi (\vec x)]$ 
and the Hamiltonian of the system by ${\cal H}[\varphi (\vec x)]$.
The variable $\varphi$ may be a scalar or a vector
and $\vec x$ may denote either the points in a continous region or the
sites of a lattice. The set of all possible configurations forms the phase
space $\Omega$. Symmetries of the system are the groups of transformations
of $\Omega$ onto itself, which leave ${\cal H}[\varphi (\vec x)]$ invariant.
The equilibrium state at the temperature $T$ is the Gibbs state
\be
P_{G}[\varphi (\vec x)] = {1 \over Z}\exp \{- {\cal H}[\varphi (\vec x)]/T \}
\label{2.3}
\ee
where $Z= \sum_{[\varphi (\vec x)]}\exp \{- {\cal H}[\varphi (\vec x)]/T \}$
and the Boltzmann constant is taken $k_B=1$. It is evident that 
$P_{G}[\varphi (\vec x)]$ shares the symmetries of the Hamiltonian. 

The models considered are characterized by the existence of a critical temperature $T_C$ and
a phase diagram as in Fig.\ref{fig1}.
For $T \geq T_C$ the symmetry is not broken and $P_{G}[\varphi (\vec x)]$ is a pure state.
For  $T < T_C$ the symmetry is broken and $P_{G}[\varphi (\vec x)]$ is a mixture  
\be
P_{G}[\varphi (\vec x)] = \sum_{\alpha} w_{\alpha} P_{\alpha}[\varphi (\vec x)] 
\label{2.4}
\ee
where the broken symmetry pure states $P_{\alpha}[\varphi (\vec x)]$ are invariant
under a subgroup of the symmetry group of the Hamiltonian, while the weights $ w_{\alpha} \geq 0$ 
satisfy $\sum_{\alpha}w_{\alpha} =1$. In the Gibbs state,
due to symmetry, all the weights are equal
\be
w_{\alpha} = w, \qquad \forall \; \alpha. 
\label{2.4bis}
\ee
Thus, for example, in the case of the Ising model, where the order parameter configurations
are spin configurations $[s_i=\pm 1]$ on a lattice, below $T_C$ there are two pure states, labeled by $\alpha=\pm$,
which transform one into the other under spin inversion
\be
P_+[s_i]=P_-[-s_i]
\label{2.00}
\ee
and remain invariant under the trivial subgroup of the identity.
The Gibbs state is given by
\be
P_{G}[s_i]= w_+P_+[s_i] +  w_-P_-[s_i] 
\label{2.01}
\ee
with $w_+=w_-=w=1/2$.
With a vector order parameter and ${\cal H}[\vec \varphi(\vec x)]$ invariant under rotations, 
the labels of the pure states are the unit vectors $\hat {\alpha}$ in the
the order parameter space and each broken symmetry state 
$P_{\hat {\alpha}}[\vec \varphi(\vec x)]$ is invariant under the subgroup of the rotations around  
the $\hat {\alpha}$ axis.

\subsection{Magnetization}

We shall assume throughout, except when it will be explicitely stated otherwise,
that all expectation values are space translation invariant. Then, the equilibrium magnetization
is given by 
\be
m_{eq}= \langle \varphi (\vec x) \rangle_{G} 
\label{2.5}
\ee
where $\langle \cdot \rangle_{G}$ denotes averages taken with respect to $P_{G}[\varphi (\vec x)]$.
Since $\varphi (\vec x)$ is not symmetrical,  
$m_{eq}$ must vanish for all temperatures.
Namely, the existence of the phase transition cannot be
detected from the magnetization in the Gibbs state. 
Below  $T_C$, using Eqs.~(\ref{2.4}) and~(\ref{2.4bis})
\be
m_{eq}=w \sum_{\alpha} m_{\alpha} =0
\label{2.6}
\ee
where $m_{\alpha} = \langle \varphi (\vec x) \rangle_{\alpha} \neq 0$ is 
the spontaneous magnetization in the pure state $P_{\alpha}[\varphi (\vec x)]$.
What vanishes below $T_C$ is the sum
of all the possible values of the spontaneous magnetization, 
but separately each contribution is not zero.

\subsection{Correlation function}

Conversely, the behavior of the order parameter correlation function
\be
C_{eq}(\vec r,T) = \langle  \varphi (\vec x+ \vec r) \varphi (\vec x) \rangle_{G}
-  \langle \varphi (\vec x+ \vec r) \rangle_{G}  \langle \varphi (\vec x) \rangle_{G}
\label{2.8}
\ee
allows to detect the existence of the
phase transition, even in the symmetrical Gibbs state.
From the scaling behavior for $T \geq T_C$
\be
C_{eq}(\vec r,T) = r^{-(d-2+\eta)} f_{eq}(r/\xi)
\label{2.9}
\ee
where $\xi$ is the correlation length and $f_{eq}(x)$ is a rapidly vanishing scaling function, 
follows the clustering property 
\be
\lim_{r \to \infty} C_{eq}(\vec r,T) = 0.
\label{2.9tris}
\ee
Instead, for $T < T_C$, from Eqs.~(\ref{2.8}) and~(\ref{2.4}) follows
\be
C_{eq}(\vec r,T) = \sum_{\alpha} w_{\alpha} C_{\alpha}(\vec r,T) + q_{eq}
\label{2.11}
\ee
where
\be
C_{\alpha}(\vec r,T) =  \langle  \varphi (\vec x+ \vec r) \varphi (\vec x) \rangle_{\alpha}
- \langle \varphi (\vec x+ \vec r) \rangle_{\alpha} \langle \varphi (\vec x) \rangle_{\alpha}  
\label{2.12}
\ee
is the correlation function in the $\alpha$-th broken symmetry state
and 
\be
q_{eq}=\sum_{\alpha} w_{\alpha} m_{\alpha}^2 - \left [\sum_{\alpha} w_{\alpha} m_{\alpha} \right ]^2
\label{2.13}
\ee
is the variance of the spontaneous magnetization in the Gibbs state.
This quantity in the spin glass context is the Edwards-Anderson order parameter~\cite{MPV}.
Since $C_{\alpha}(\vec r,T)$ and $m_{\alpha}^2$ are independent 
of $\alpha$, it is convenient to introduce the notation 
$ G_{eq}(\vec r,T)=C_{\alpha}(\vec r,T)$, $M^2=m_{\alpha}^2$
and to rewrite 
\be
C_{eq}(\vec r,T) = G_{eq}(\vec r,T) + M^2
\label{2.14}
\ee
where $G_{eq}(\vec r,T)$ has a form similar to~(\ref{2.9}) 
\be
G_{eq}(\vec r,T) =  r^{-(d-2+\eta)} g_{eq}(r/\xi). 
\label{2.15}
\ee
The appearence of a non zero value of $M^2$ upon crossing $T_C$ 
\be
\lim_{r \to \infty}C_{eq}(\vec r,T) = M^2 \neq 0
\label{2.15}
\ee
signals the occurrence of the phase transition and the breaking of ergodicity~\cite{Palmer}. 
Notice that  $C_{eq}(\vec r,T)$ is a connected correlation function and its decay
to a non vanishing value at large distances implies that the correlation
length $\xi_{G}$ in the Gibbs state is
divergent for $T < T_C$, as opposed to $\xi$ in the pure states, which is
finite for $T <T_C$.

\subsection{Splitting of the order parameter}

The structure~(\ref{2.14}) of the correlation function can be viewed as due to
the splitting of the order parameter into the sum of two statistically independent
contributions
\be
\varphi (\vec x)= \psi (\vec x) + \sigma
\label{2.20}
\ee
each with zero mean
\be
\langle \psi (\vec x) \rangle = \langle \sigma \rangle = 0
\label{2.21}
\ee
and such that
\be
G_{eq}(\vec r,T) = \langle \psi (\vec x+ \vec r) \psi (\vec x) \rangle
\label{2.22}
\ee
\be
M^2= \langle \sigma^2 \rangle.
\label{2.23}
\ee
The first contribution represents the thermal fluctuations in any of the pure states 
and is obtained by shifting the order parameter by its mean
\be
\psi (\vec x)= \varphi (\vec x) - m_{\alpha}.
\label{2.16}
\ee
This quantity averages to zero by construction and the probability distribution 
\be
P[\psi (\vec x)]=P_{\alpha}[\psi (\vec x) + m_{\alpha}]
\label{2.17}
\ee
is independent of $\alpha$,
since the deviations from the mean are equally distributed in all pure states.
The second contribution $\sigma$ fluctuates over the possible values of the spontaneous magnetization,
taking the values $m_{\alpha}$ with probabilities $P(m_{\alpha}) = w_{\alpha}$. Then, from Eqs.~(\ref{2.6})
and~(\ref{2.13}) follows
\be
\langle \sigma \rangle =0, \qquad M^2= \langle \sigma^2 \rangle.
\label{2.17bis}
\ee

\subsection{Static susceptibility}

The introduction of a field $h(\vec x)$ conjugate to the order parameter modifies
the Hamiltonian 
\be
{\cal H}_h[\varphi (\vec x)] = {\cal H}[\varphi (\vec x)] - \sum_{\vec x} h(\vec x)\varphi (\vec x)
\label{2.25}
\ee
and the corresponding Gibbs state 
\be
P_{G,h}[\varphi (\vec x)] = {1 \over Z_h}\exp \{ -{\cal H}_h[\varphi (\vec x)]/T \}.
\label{2.26}
\ee
With an $\vec x$ dependent external field averages are no more space translation invariant and,
if the field is small, the magnetization at the site $\vec x$ is given by
\be
\langle \varphi (\vec x) \rangle_{G,h} = \langle \varphi (\vec x) \rangle_0
+ \int d\vec y \; \chi_{st}(\vec x-\vec y,T) h(\vec y) + {\cal O}(h^2)
\label{2.27}
\ee
where $\langle \varphi (\vec x) \rangle_0$ is the magnetization in the state
\be
   P_0[\varphi (\vec x)] = \lim_{h \to 0} P_{G,h}[\varphi (\vec x)] = \left \{ \begin{array}{ll}
        P_{G}[\varphi (\vec x)], \qquad $for$ \qquad T \geq T_C \\
        P_{\alpha}[\varphi (\vec x)],  \qquad $for$ \qquad T < T_C
        \end{array}
        \right .
        \label{p.7}
        \ee
and $P_{\alpha}[\varphi (\vec x)]$ here stands for the particular
broken simmetry state selected when the external
field is switched off. The site dependent static susceptibility is given by
\be
\chi_{st}(\vec x-\vec y,T) = \left . {\delta \langle \varphi (\vec x) \rangle_{G,h} 
\over \delta h(\vec y)} \right |_{h=0}
\label{2.28}
\ee
and making the expansion, up to first order in $h(\vec x)$, also 
in the definition
\be
\langle \varphi (\vec x) \rangle_{G,h} = \sum_{[\varphi (\vec x^{\prime})]}\varphi (\vec x) 
P_{G,h}[\varphi (\vec x^{\prime})]
\label{2.280}
\ee
from the comparison with Eq.~(\ref{2.27}) it is straightforward to derive the 
fluctuation-response theorem
\be
\chi_{st}(\vec r,T) = {1 \over T} C_{0}(\vec r,T)
\label{2.281}
\ee
where
$C_{0}(\vec r,T)$ is the correlation function in the state $P_0[\varphi (\vec x)]$.

\section{Dynamics}

Let us now examine the time dependent properties in 
the various quenches described in the introductory section.
The order parameter expectation value and correlator in the infinite temperature initial 
state $P_G[\varphi,T_I]$ are given by
\be
\langle \varphi (\vec r) \rangle_I=0
\label{5.1}
\ee 
\be
\langle \varphi (\vec r)\varphi (\vec r^{\prime}) \rangle_I= \Delta \delta(\vec r -\vec r^{\prime}). 
\label{5.2}
\ee
Modelling the dynamics with a Markov stochastic process, the time dependent probality distribution
evolves with the master equation
\be
{\partial \over \partial t}P[\varphi,t] = \sum_{[\varphi^{\prime}]}
\Gamma([\varphi]|[\varphi^{\prime}])  P[\varphi^{\prime},t]
\label{5.3}
\ee
where $\Gamma([\varphi]|[\varphi^{\prime}])$ is the transition probability per unit time
from $[\varphi^{\prime}]$ to $[\varphi]$,
satisfying the detailed balance condition with the Gibbs state at the final temperature $T_F$
\be
\Gamma([\varphi]|[\varphi^{\prime}]) P_G[\varphi^{\prime},T_F]=
\Gamma([\varphi^{\prime}]|[\varphi]) 
P_G[\varphi ,T_F].
\label{5.4}
\ee
Therefore, the dynamics preserves the symmetry of the Hamiltonian and, {\it if} equilibrium is reached,
then {\it necessarily} $P[\varphi,t]$ must go over to $P_G[\varphi,T_F]$. The crucial question, of course,
is whether the system equilibrates or not. As anticipated in the Introduction,
four qualitatively different relaxation processes arise, depending on the values of $d$ and $T_F$

\begin{enumerate}

\item quench to $T_F > T_C$: there is a finite equilibration time $t_{eq}$

\item quench to $T_F = T_C$: equilibrium is reached in an infinite time

\item quench to $(d>d_L,T_F < T_C)$: $C(t,t_w)$ does not equilibrate, while the
dynamic susceptibility $\chi(t,t_w)$ equilibrates

\item quench to $(d=d_L,T_F=0)$: neither $C(t,t_w)$ nor $\chi(t,t_w)$ do equilibrate.

\end{enumerate}

\noindent The dynamic susceptibility $\chi(t,t_w)$ will be defined more precisely in section~\ref{LRF}
as the zero field cooled susceptibility.

\subsection{Equilibration}

Since $P[\varphi,t]$ remains symmetrical during the relaxation, no information
on the equilibration process can be extracted from the time dependent magnetization,
which vanishes throughout $m(t) = \langle\varphi (\vec r,t) \rangle \equiv 0$. 
One must turn to the time dependent correlation function
\be
C(\vec r,t,t_w) =  \langle \varphi (\vec x+ \vec r,t) \varphi (\vec x,t_w)\rangle
- \langle \varphi (\vec x+ \vec r,t) \rangle \langle\varphi (\vec x,t_w)\rangle
\label{5.5}
\ee
where the angular brackets denote the average over the initial condition and the thermal
noise. For what is needed in the following
it is sufficient to consider the autocorrelation function
$C(t,t_w) = C(\vec r=0,t,t_w)$.

If there exists a finite equilibration time $t_{eq}$, then for $t_w > t_{eq}$
the dynamics becomes TTI with
\be
C(t,t_w) = C_{eq}(\tau,T_F).
\label{5.7}
\ee
So, if it is not known how large $t_{eq}$ is, or if it exists at all, in order to ascertain
whether equilibration has occurred or not, it is necessary to look at the large $t_w$ behavior.
The $t_w \rightarrow \infty$ limit requires to specify also how $t$ is pushed to infinity.
This is done by rewriting $C(t,t_w)$ in terms of the new pairs of variables $(\tau,t_w)$ and
$(x=t/t_w,t_w)$
\be
C(t,t_w) = \widetilde{C}(\tau,t_w) = \widehat{C}(x,t_w)
\label{5.70}
\ee
and, then, by taking the limit $t_w \rightarrow \infty$ while keeping either $\tau$ or $x$ fixed. 
The short $\tau \ll t_w$ and the large $\tau \gg t_w$ time separation
regimes are explored, respectively, in the first and
in the second case. Notice that from $x=\tau/t_w + 1$ follows that when using the $x$ variable the
short time regime gets all compressed into $x=1$.
Assuming that the limits exist, one has
\be
\lim_{t_w \to \infty}\widetilde{C}(\tau,t_w) = K_C(\tau)
\label{5.71}
\ee
and 
\be
\lim_{t_w \to \infty}\widehat{C}(x,t_w) ={\cal C}(x)
\label{5.72}
\ee
which, in general, are two functions not related one to the other.
However, if there exists an equilibration time  $t_{eq}$ and Eq.(\ref{5.7}) holds on all time scales, 
then $K_C(\tau)= C_{eq}(\tau,T_F)$ and for $t_w > t_{eq}$ 
\be
\widehat{C}(x,t_w) = C_{eq}((x-1)t_w,T_F)
\label{5.75}
\ee
which implies the singular limit
\be
   {\cal C}(x)  = \left \{ \begin{array}{ll}
        C_{eq}(0,T_F), \qquad $for$ \qquad x=1 \\
        C_{eq}(\infty,T_F),  \qquad $for$ \qquad x>1.
        \end{array}
        \right .
        \label{5.13}
        \ee
This is a necessary condition for equilibration. To be also a sufficient condition it should hold for all
two times observables. As we shall see below it may hold for some, but not for others.

\subsection{Generic properties of $C(t,t_w)$}
\label{generic-C}

In the quench to $T_F > T_C$, since there is a finite equilibration time, Eq.~(\ref{5.13})
necessarily holds. The interesting cases are those of the quenches to and to below $T_C$, where
$t_{eq}$ diverges. 

In the quench to $T_C$, the generic form of $C(t,t_w)$ displays the multiplicative combination of 
the stationary and aging contributions~\cite{Godreche02,Calabrese}
 \be
C(t,t_w)=(\tau + t_0)^{-b_c}g_C(t/t_w)
\label{5.14}
\ee
where 
\be
b_c=(d-2+\eta)/z_c
\label{5.15}
\ee
$\eta$ is the usual static exponent, $z_c$ is the dynamical critical exponent~\cite{HH} and
$t_0$ is a microscopic time\footnote{For instance, $t_0$ might be the time it takes
for a domain wall to advance by one lattice spacing.}
needed to regularise the equal time autocorrelation function. 
Taking the short time limit~(\ref{5.71}) 
$K_C(\tau)$ is found to coincide with the autocorrelation function of equilibrium critical dynamics  
\be
C_{eq}(\tau,T_C)=(\tau + t_0)^{-b_c}g_{C}(1)
\label{5.17}
\ee
where $C_{eq}(0,T_C)= t_0^{-b_c}g_{C}(1) = \langle \varphi^2(\vec x) \rangle_G$.
In order to explore the
large time regime,
Eq.~(\ref{5.14}) can be rewritten in the simple aging form
\be
C(t,t_w)= t_w^{-b_c}f_C(x,y)
\label{5.140}
\ee
with $y=t_0/t_w$. The scaling function 
\be
f_C(x,y) = (x-1+y)^{-b_c}g_C(x)
\label{5.141}
\ee
decreases asymptotically with the power law
\be
f_C(x,y) = A_C x^{-\lambda_c/z_c}
\label{5.16}
\ee
where $\lambda_c$ is the critical autocorrelation exponent~\cite{Huse}. Then, taking the
$t_w \rightarrow \infty$ limit one finds
\be
   {\cal C}(x)  = \left \{ \begin{array}{ll}
        t_0^{-b_c}g_C(1), \qquad $for$ \qquad x=1 \\
        0,  \qquad $for$ \qquad x>1
        \end{array}
        \right .
        \label{5.190}
        \ee
in agreement with Eq.~(\ref{5.13}). Therefore, although it is clear that equilibrium is not reached in 
any finite time, the autocorrelation function behaves as if equilibrium
was reached in an infinite time.

\begin{figure}[h]
\begin{center}
\includegraphics*[scale = 0.5]{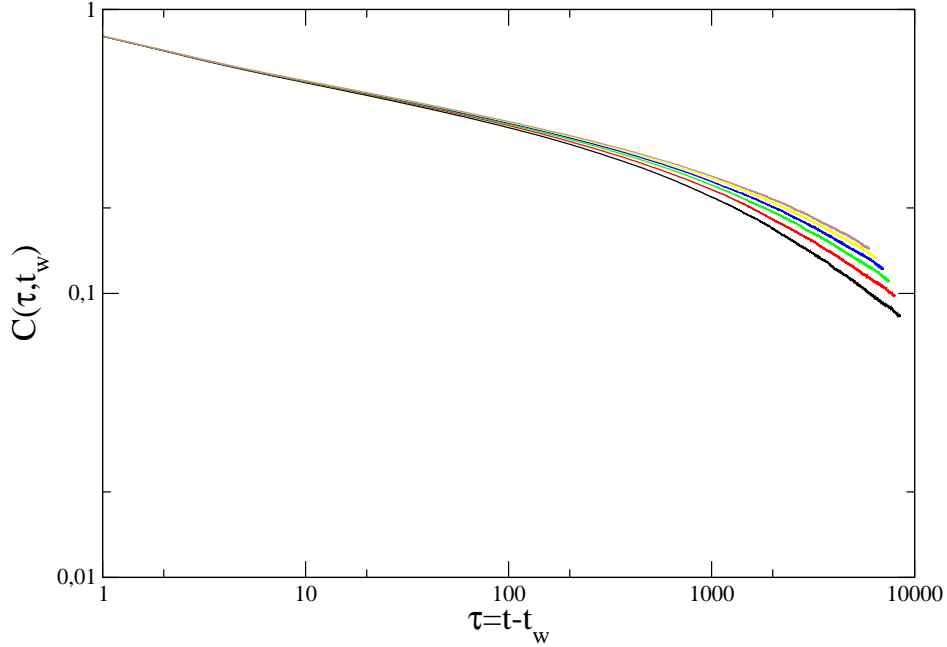}
\end{center}
\caption{\label{fig3}$\widetilde{C}(\tau,t_w)$ vs. $\tau$ in the quench to $T_C$ of the $d=2$ kinetic Ising model.
$t_w$ is incresing by $500$ from $1500$ up to $4000$ (bottom to top). In this and in the following
figures times are always in units of Monte Carlo steps.}
\end{figure}

\begin{figure}[h]
\begin{center}
\includegraphics[scale = 0.5]{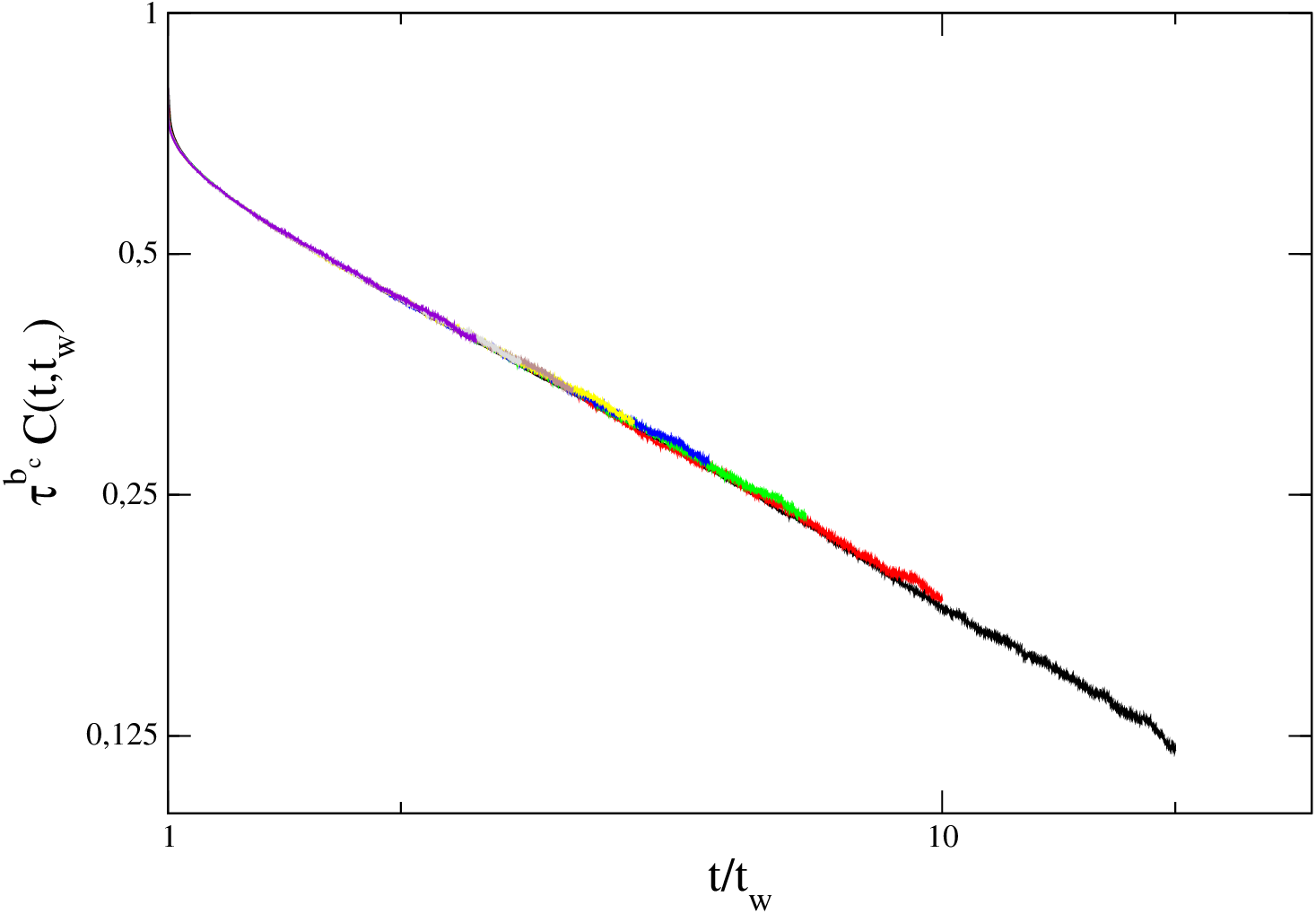}
\end{center}
\caption{\label{fig4}$\tau^{b_c}C(t,t_w)$ vs. $x=t/t_w$ in the quench to $T_C$ of the $d=2$ kinetic Ising model
and for the same values of $t_w$ as in Fig.\ref{fig3}. The collapse of the different curves demonstrates the
multiplicative structure of Eq.(\ref{5.14}).}
\end{figure}

As an illustration, the behavior of $C(t,t_w)$
in the quench to $T_C$ of the $d=2$ kinetic Ising model with Glauber dynamics~\cite{GALG} is displayed in
Figs. \ref{fig3} and~\ref{fig4}. In the first one $\widetilde{C}(\tau,t_w)$ is plotted agains $\tau$
for increasing values of $t_w$, showing
quite well the separation of the time scales, since the curves collapse in the short time
regime and spread out as the large time regime is entered. Furthermore, the collapse improves with
increasing $t_w$, showing the convergence of $\widetilde{C}(\tau,t_w)$ toward $C_{eq}(\tau,T_C)$.
The multiplicative structure is demonstrated in Fig.\ref{fig4}, where the plot of $\tau^{b_c}C(t,t_w)$
against $x$ for different values of $t_w$ shows the collapse 
of the data as required by Eq.~(\ref{5.14}), since in the chosen $t_w$ range 
$t_0$ is negligible. 
This plot has been made using $b_c=0.115$  
obtained from Eq.~(\ref{5.15}) with $\eta=1/4$ and $z_c=2.167$~\cite{wang}.

\begin{figure}[h]
\begin{center}
\includegraphics*[scale = 0.5]{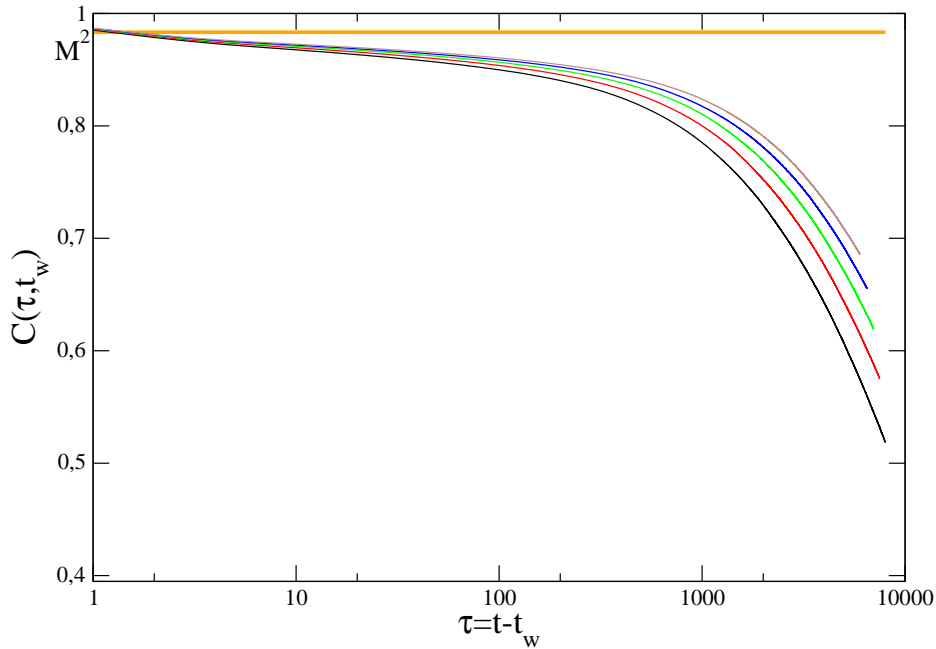}
\end{center}
\caption{\label{fig5}$\widetilde{C}(\tau,t_w)$ vs. $\tau$ in the quench to $T_F = 0.66 \; T_C$ of 
the $d=2$ kinetic Ising model.
$t_w$ is incresing by $500$ from $2000$ up to $4000$ (bottom to top). The orizontal line indicates the
Edwards-Anderson order parameter $M^2=0.97$.}
\end{figure}

\begin{figure}[h]
\begin{center}
\includegraphics*[scale = 0.5]{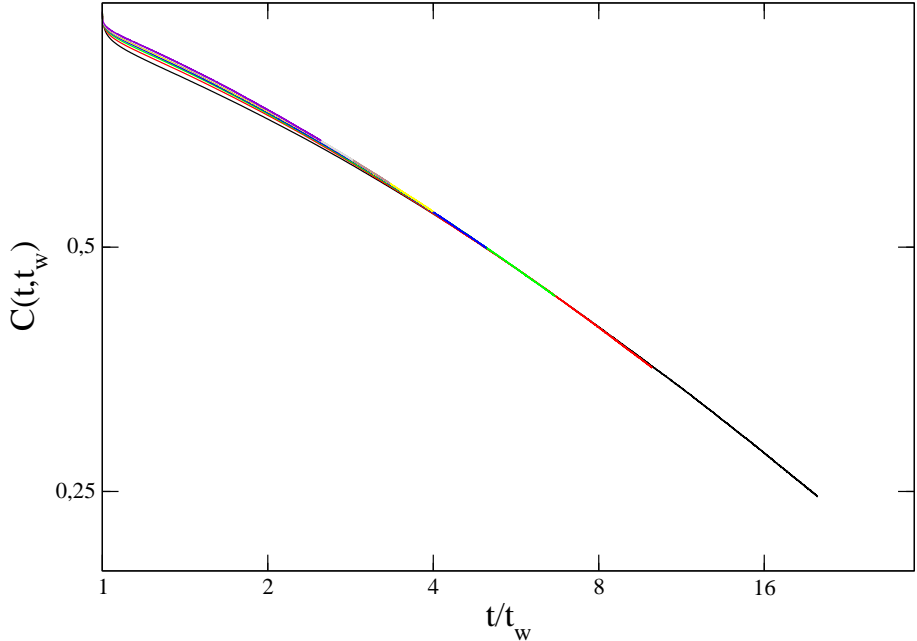}
\end{center}
\caption{\label{fig6}$C(t,t_w)$ vs. $x=t/t_w$ in the quench to $T_F = 0.66 \; T_C$ of the $d=2$ kinetic Ising model
and for the same values of $t_w$ as in Fig.\ref{fig5}.}
\end{figure}

When the quench is made to $T_F < T_C$, in the short time regime the $t_w \rightarrow \infty$ limit
gives, again, the equilibrium behavior
\be
K_C(\tau) = C_{eq}(\tau,T_F)=  G_{eq}(\tau,T_F) + M^2.
\label{K.1}
\ee
For large time, instead, one finds
\be
    {\cal C}(x) = \left \{ \begin{array}{ll}
        C_{eq}(0,T_F), \qquad $for$ \qquad x=1 \\
        h_C(x),  \qquad $for$ \qquad x>1
        \end{array}
        \right .
        \label{5.22}
        \ee
where $h_C(x)$ is a monotonously decaying function with the limiting behaviors
\be
     h_C(x) = \left \{ \begin{array}{ll}
        M^2, \qquad $for$ \qquad x=1 \\
        x^{-\lambda/z},  \qquad $for$ \qquad x \gg 1
        \end{array}
        \right .
        \label{5.23}
        \ee
and $\lambda$ is the autocorrelation exponent below $T_C$~\cite{Fisher}.
Therefore, i) the necessary condition~(\ref{5.13}) for equilibration is violated and ii) the
above result, together with Eq.~(\ref{K.1}), implies the non commutativity of the limits 
$t_w \rightarrow \infty$ and $t \rightarrow \infty$ with
\be
\lim_{\tau \to \infty} \lim_{t_w \to \infty} C(\tau + t_w,t_w)= M^2
\label{web}
\ee
and
\be
\lim_{t_w \to \infty} \lim_{t \to \infty} C(t,t_w)= 0.
\label{web.1}
\ee
This is the phenomenon of weak ergodicity
breaking~\cite{WEB}, since ergodicity appears broken in the short time regime,
but not in that of the large time separations. The behaviors described
above are illustrated in Figs. \ref{fig5} and~\ref{fig6}, where 
$C(t,t_w)$, computed in the quench to $T_F=0.66\; T_C$
of the $d=2$ kinetic Ising model, is
plotted against $\tau$ and $x$. Both plots highlight the
separation of time scales, with the collapse of the data either in the short or in the large
time regime, as well as the falling of the curves below the Edwards-Anderson plateau at $M^2=0.97$,
revealing that the system keeps on decorrelating for arbitrarily large time scales. 
The curve in Fig.\ref{fig6} for $x>3$, where the collapse is good, essentially
is a plot of $h_C(x)$.

Weak ergodicity breaking is incompatible with the multiplicative form~(\ref{5.14}). Then,
in order to put together the separation of the time scales and weak ergodicity breaking,
$C(t,t_w)$ must have the additive structure
\be
C(t,t_w) = G_{eq}(\tau,T_F)  + C_{ag}(t,t_w) 
\label{6.1}
\ee
where $C_{ag}(t,t_w)$ obeys the simple aging form~(\ref{I.2}) with the scaling function $h_C(x)$ and $b=0$. 
Notice that this value of $b$ is of geometrical origin, since it is related through the
multitime scaling of Furukawa~\cite{Furukawa} 
\be
C_{ag}(\vec r,t,t_w)= {\cal F}(r/t^{1/z},t/t_w)
\label{furukawa}
\ee
to the well known~\cite{Bray94} scaling of the equal time correlation function
\be
C_{ag}(\vec r,t)= F(r/t^{1/z})= {\cal F}(r/t^{1/z},1)
\label{eqtime}
\ee
which, in turn, is a consequence of the compact nature of the ordered domains\footnote{In 
the ususal treatment of phase-ordering 
kinetics~\cite{Bray94} the thermal contribution $G_{eq}$
is either negligible or absent, when quenches to $T_F=0$ are considered.}.
Here $z$ is the phase-ordering growth exponent with the value $z=2$ for dynamics with 
non conserved order parameter~\cite{Bray94}.

\subsection{Splitting of the order parameter}

The additive structure~(\ref{6.1}) leads, for large time, to the generalization of the splitting~(\ref{2.20})  
of the order parameter into the sum of two time dependent components
\be
\varphi (\vec x,t)= \psi (\vec x,t) + \sigma(\vec x,t)
\label{6.4}
\ee
in such a way that the two contributions are still statistically independent, of zero mean and with the
respective autocorrelation functions given by
\be
\langle \psi (\vec x,t) \psi (\vec x,t_w) \rangle = G_{eq}(\tau,T_F)
\label{6.5}
\ee
\be
\langle \sigma (\vec x,t) \sigma (\vec x,t_w) \rangle = C_{ag}(t,t_w).
\label{6.6}
\ee
The variables $\psi (\vec x,t)$ and $\sigma(\vec x,t)$ are associated 
to the fast and slow degrees of freedom~\cite{Mazenko88,Franz,EPJ}. 
Keeping in mind the domain structure of the
configurations, the local ordering variable is defined by
\be
\sigma(\vec x,t) = m_{\alpha(\vec x,t)}
\label{6.7}
\ee
where $\alpha(\vec x,t)$ is the value of $\alpha$ selected by the domain to which the site $\vec x$
belongs at the time $t$. 
Hence, $\sigma(\vec x,t)$ represents the local equilibrium magnetization
within domains, while $\psi (\vec x,t)$ is the field of thermal fluctuations, as in
Eq.~(\ref{2.20}).
The off-equilibrium character of the dynamics
enters only through the ordering variable $\sigma(\vec x,t)$. In fact, while $\psi (\vec x,t)$
executes the equilibrium thermal fluctuations, $\sigma(\vec x,t)$ can change only if a defect,
or an interface, goes by the site $\vec x$ at the time $t$. In other words, the evolution
of $\sigma(\vec x,t)$ is strictly related to the existence of defects in the system, which is
precisely what keeps the system out of equilibrium.
As we shall see in section~\ref{largeN}, in the case of the large $N$ model the construction~(\ref{6.4}) 
can be carried out exactly.

\section{Linear response function}
\label{LRF}

As previously stated, the quench of phase-ordering systems offers the full spectrum of off-equilibrium
phenomena, with increasing degree of deviation from equilibrium
as $T_F$ is lowered from above to below $T_C$. The survey of aging, so far, 
has been conducted using the order parameter autocorrelation
function as a probe. However, once the basic features of the phenomena involved have been brought
into focus, in order to carry out a more refined analysis and to make progress
in the characterization of the deviation from equilibrium, it is indispensable
to look jointly at the autocorrelation and 
autoresponse function. In particular, the deviations from the
fluctuation-dissipation theorem (FDT) have proven to be a most effective tool of 
investigation~\cite{Bouchaud,CugliandoloLH}.
This approach to the study of out of equilibrium dynamics has been pionereed by
Cugliandolo and Kurchan in their groundbreaking work on the
mean field models of the spin glass~\cite{CK}.

Let us begin by defining the time dependent linear response function.
If a small space and time dependent external field $h(\vec x,t)$
is switched on in the time interval $(t_1,t_2)$ after the quench, then 
the magnetization at the time $t \geq t_2$ is
given by
\be
\langle \varphi (\vec x,t) \rangle_h = \langle \varphi (\vec x,t) \rangle +
\int d\vec y \;\int_{t_1}^{t_2}dt^{\prime}\; R(\vec x-\vec y,t,t^{\prime})h(\vec y,t^{\prime}) + {\cal O}(h^2)
\label{3.010}
\ee
where $\langle \varphi (\vec x,t) \rangle$ is the magnetization in the absence of the field and
\be
R(\vec x-\vec y,t,t_w) = {\left . \delta \langle \varphi (\vec x,t) \rangle_h
\over \delta h(\vec y, t_w) \right |_{h=0}}
\label{3.02}
\ee 
is the space and time dependent linear response function. The autoresponse function $R(t,t_w)$
is obtained taking $\vec x = \vec y$.

With a time independent external field, Eq.~(\ref{3.010}) takes the form
\be
\langle \varphi (\vec x,t) \rangle_h = \langle \varphi (\vec x,t) \rangle +
\int d\vec y \;\zeta(\vec x-\vec y,t,t_1,t_2)h(\vec y) + {\cal O}(h^2)
\label{3.020}
\ee 
where
\be
\zeta(\vec r,t,t_2,t_1) = 
\int_{t_1}^{t_2}dt^{\prime} R(\vec r,t,t^{\prime})
\label{3.021}
\ee
is the integrated linear response function. Particular cases, frequently encountered in the 
literature, are those of the thermoremanent magnetization (TRM) corresponding to the
protocol $t_1=0$, $t_2=t_w$
\be
\rho(\vec r,t,t_w) = \zeta(\vec r,t,t_w,0)
\label{3.022}
\ee
and of the zero field cooled (ZFC) susceptibility corresponding to $t_1=t_w$, $t_2=t$
\be
\chi(\vec r,t,t_w) = \zeta(\vec r,t,t,t_w).
\label{3.023}
\ee

\subsection{Fluctuation-dissipation theorem (FDT)}

For convenience, let us briefly derive the FDT. Assuming that the small external field
has been applied from a time so distant in the past that equilibrium in the field is established
at the time $t_w$ and that it is switched off for $t > t_w$, the magnetization 
is given by
\be
\langle \varphi (\vec x,t) \rangle_h =
\sum_{[\varphi],[\varphi^{\prime}]}\varphi(\vec x)Q([\varphi,t]|[\varphi^{\prime},t_w])
P_{G,h}[\varphi^{\prime}]
\label{7.1}
\ee
where $Q([\varphi,t]|[\varphi^{\prime},t_w])$ is
the conditional probability in the absence of the field, since $t>t_w$. Recalling that $P_{G,h}[\varphi]$
is given by Eqs.~(\ref{2.25}) and~(\ref{2.26}) and expanding up to first order in the field, one finds
\be
\langle \varphi (\vec x,t) \rangle_h = \langle \varphi (\vec x,t) \rangle +{1 \over T_F}
\int d\vec y \;C_{0}(\vec x-\vec y,t-t_w,T_F)h(\vec y)
\label{7.2}
\ee
where $C_{0}(\vec x-\vec y,t-t_w,T_F)$ is the equilibrium, unperturbed correlation function
in the stationary state~(\ref{p.7}). Hence, comparing with Eq.~(\ref{3.020})
\be
{1 \over T_F}C_{0}(\vec r,t-t_w,T_F)= \int_{-\infty}^{t_w} dt^{\prime} R(\vec r,t,t^{\prime})
\label{7.2a}
\ee
and differentiating with respect to $t_w$ the FDT is obtained
\be
R_{eq}(\vec r,\tau,T_F) = -{1 \over T_F}{\partial \over \partial \tau}C_{0}(\vec r,\tau,T_F)
\label{7.4}
\ee
where $R_{eq}(\vec r,\tau,T_F)$ is the equilibrium response function.
From this, it is straightforward to derive the integrated form of the FDT in terms of
the equilibrium ZFC susceptibility
\be
\chi_{eq}(\vec r,\tau,T_F) =  {1 \over T_F}[C_{0}(\vec r,T_F)  - C_{0}(\vec r,\tau,T_F)]
\label{7.5}
\ee
and using $\lim_{\tau \to \infty}C_{0}(\vec r,\tau,T_F)=0$, this gives
the identification, via Eq.~(\ref{2.281}), of the large time limit
of the equilibrium ZFC susceptibility with the static susceptibility
\be
\lim_{\tau \to \infty} \chi_{eq}(\vec r,\tau,T_F) =  \chi_{st}(\vec r,T_F).
\label{3.024}
\ee

\subsection{Generic properties of $R(t,t_w)$}
\label{generic-R}

Before exploring the deviations from the FDT when the system is not in equilibrium, 
it is convenient to go over the
generic properties of $R(t,t_w)$, as it has been done for $C(t,t_w)$ in section~\ref{generic-C}. 
Apart for the few cases where
analytical results are available, $R(t,t_w)$ is less known 
than $C(t,t_w)$, since it is much more difficult to measure
numerically. Actually, untill very recently, $R(t,t_w)$ was numerically
accessible only indirectly through the measurement of the integrated response functions.
This situation has partially changed after the introduction of new and 
more efficient algorithms~\cite{Chat1,noialg,corrsc}.
In any case, whenever TTI holds,
like in equilibrium or in the short time sector, the task is easier
since the form of $R(t,t_w)$ is related to that of $C(t,t_w)$ via
the FDT. When TTI does not hold, scaling arguments will be used. 

In the quench to $T_C$, the analogue of the multiplicative form~(\ref{5.14}) reads
\be
R(t,t_w)=(\tau + t_0)^{-(1+a_c)}g_R(t/t_w)
\label{8.1}
\ee
whose short time limit $K_R(\tau)$ coincides with the equilibrium response function
\be
R_{eq}(\tau,T_C)=(\tau + t_0)^{-(1+a_c)}g_R(1)
\label{8.6}
\ee
after requiring to be related to $C_{eq}(\tau,T_C)$ by the FDT. This yields the constraints
\be
a_c=b_c
\label{8.3}
\ee
and
\be
T_C g_R(1) = b_c g_C(1).
\label{8.6}
\ee
Switching to the $(x,t_w)$ variables, $R(t,t_w)$ is rewritten in the simple
aging form  
\be
R(t,t_w)= t_w^{-(1+a_c)}f_R(x,y)
\label{8.100}
\ee
where the scaling function 
\be
f_R(x,y) = (x-1+y)^{-(1+a_c)}g_R(x)
\label{8.102}
\ee
for large $x$ decreases with the same power law as $f_C(x,y)$~\cite{Janssen}
\be
f_R(x,y) = A_R x^{-\lambda_c/z_c}.
\label{8.101}
\ee
Then, taking the
$t_w \rightarrow \infty$ limit, the analogue of Eq.~(\ref{5.190}) is obtained
\be
   {\cal R}(x)  = \left \{ \begin{array}{ll}
        t_0^{-(1+b_c)}g_R(1), \qquad $for$ \qquad x=1 \\
        0,  \qquad $for$ \qquad x>1
        \end{array}
        \right .
        \label{5.103}
        \ee
which satisfies the necessary condition for 
equilibration
\be
   {\cal R}(x)  = \left \{ \begin{array}{ll}
        R_{eq}(0,T_F), \qquad $for$ \qquad x=1 \\
        R_{eq}(\infty,T_F),  \qquad $for$ \qquad x>1
        \end{array}
        \right .
        \label{5.104}
        \ee
in the same way as the autocorrelation function.

\begin{figure}[h]
\begin{center}
\includegraphics*[scale = 0.5]{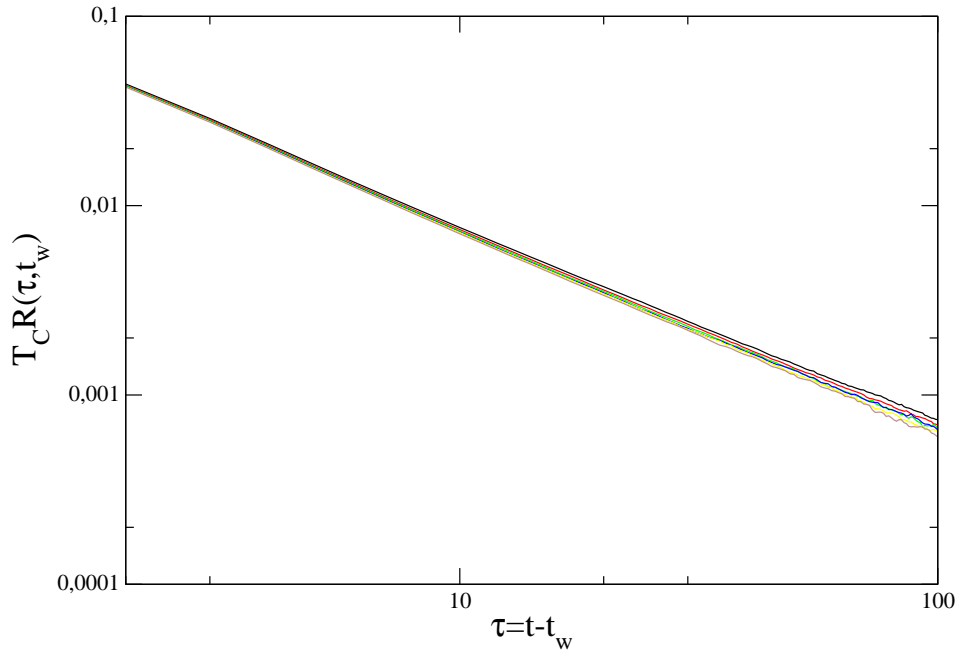}
\end{center}
\caption{\label{fig7}$T_C R(\tau,t_w)$ vs. $\tau$ in the quench to $T_C$ of the $d=2$ kinetic Ising model.
$t_w$ is incresing by $500$ from $1500$ up to $4000$ (bottom to top).}
\end{figure}

\begin{figure}[h]
\begin{center}
\includegraphics*[scale = 0.5]{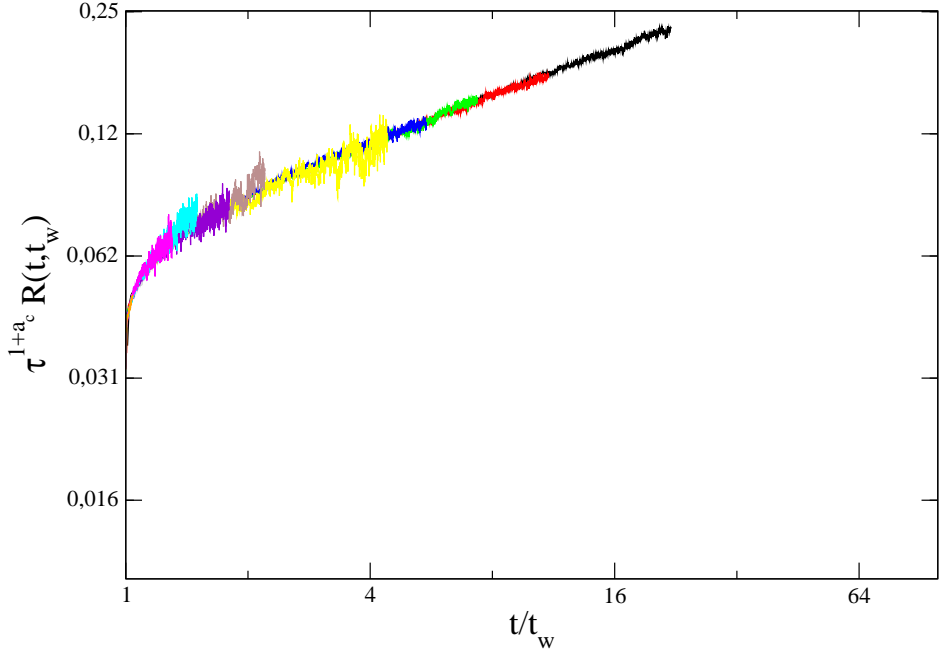}
\end{center}
\caption{\label{fig8}$\tau^{1+a_c}R(t,t_w)$ vs. $x=t/t_w$ in the quench to $T_C$ of the $d=2$ kinetic Ising model
and for the same values of $t_w$ as in Fig.\ref{fig7}. The collapse of the different curves demonstrates the
multiplicative structure of Eq.(\ref{8.1}).}
\end{figure}

\begin{figure}[h]
\begin{center}
\includegraphics*[scale = 0.5]{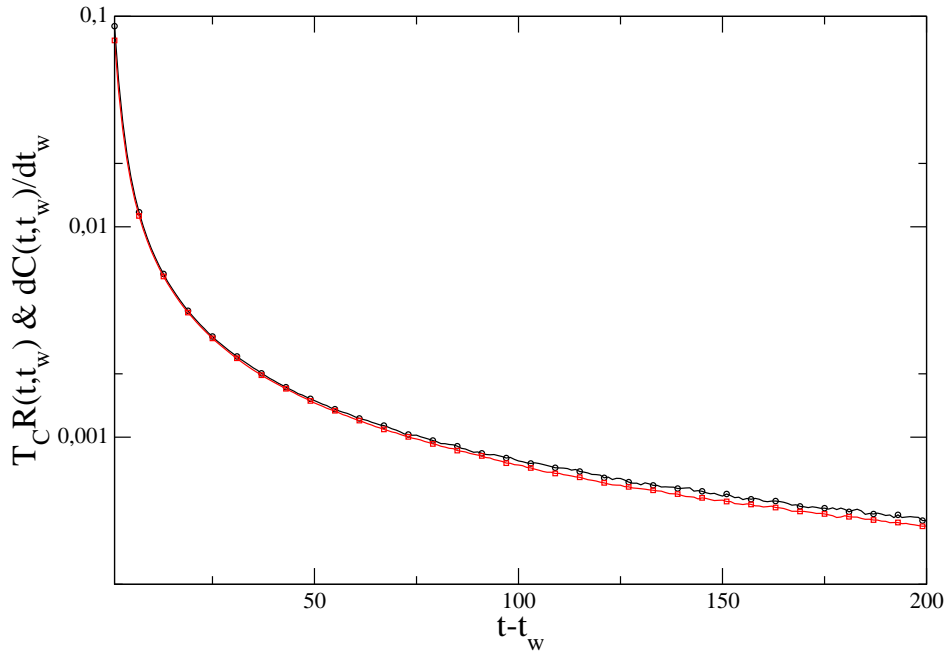}
\end{center}
\caption{\label{fig9} Plot of $T_C R(t,t_w)$ and $\partial C(t,t_w)/\partial t_w$ in the quench 
to $T_C$ of the $d=2$ kinetic Ising model, for $t_w=1000$.
The superposition of the curves for $\tau \ll t_w$ demonstrates the validity of the FDT in the
short time region.}
\end{figure}

The above behavior is well illustrated by the data for $R(t,t_w)$, obtained with the algorithm
of Ref.\cite{noialg} in
the quench to $T_C$ of the $d=2$ kinetic Ising model, along the same line of what 
has been already done for $C(t,t_w)$. In Fig.\ref{fig7} $R(t,t_w)$ is plotted agains $\tau$,
displaying, as in Fig.\ref{fig3}, the separation of the time scales, with the collapse and the
spread of the curves in the short and in the large time regimes, respectively. The collapse of 
the curves in Fig.\ref{fig8},
obtained by plotting $\tau^{1+a_c}R(t,t_w)$ against $x$, demonstrates, as in Fig.\ref{fig4}, the
multiplicative structure~(\ref{8.1}). The validity of the FDT in the short time regime 
is illustrated in  Fig.\ref{fig9}.

In the quench to below $T_C$, the additivity of $C(t,t_w)$ induces the corresponding
structure in the response function
\be
R(t,t_w) = R_{eq}(\tau,T_F) + R_{ag}(t,t_w)
\label{8.7}
\ee
in the following way: the stationary response of the fast degrees of freedom 
$R_{eq}(\tau,T_F)$ is defined by requiring the FDT to hold in the short time regime
\be
R_{eq}(\tau,T_F) = -{1 \over T_F}{\partial \over \partial \tau} G_{eq}(\tau,T_F)
\label{8.8}
\ee
where $G_{eq}(\tau,T_F)$ is the stationary contribution entering Eq.~(\ref{6.1}).
The aging component associated to the slow degrees of freedom
remains defined, thereafter, by Eq.~(\ref{8.7})
as the difference $R_{ag}(t,t_w)=R(t,t_w) - R_{eq}(\tau,T_F)$ 
and obeys the simple aging form~(\ref{I.3})
where the scaling function decays asymptotically with the same power law as $h_C(x)$~\cite{Godreche02}
\be
h_R(x) \sim x^{-\lambda/z}.
\label{8.11}
\ee
Since in the short time regime
the FDT must be satisfied also by the full $R(t,t_w)$,  
this implies that $R_{ag}(t,t_w)$ vanishes for short times.
Conversely, since $R_{eq}(\tau,T_F)$ typically is a rapidly decaying function,
in the large time regime only $R_{ag}(t,t_w)$ survives. The behavior of $R(x,t_w)$ for large $t_w$,
then, is given by
\be
    R(x,t_w)= \left \{ \begin{array}{ll}
        R_{eq}(0,T_F) \qquad $for$ \qquad x=1 \\
        t_w^{-(1+a)}h_R(x)  \qquad $for$ \qquad x>1. 
        \end{array}
        \right .
        \label{8.9}
        \ee
From this it is easy to see that, for $d > d_L$, the condition~(\ref{5.104}) is satisfied,
after taking into account that $a > 0$  
(as it will be explained shortly) and that $R_{eq}(\infty,T_F)=0$.
Therefore, contrary to what happens with the autocorrelation function, 
the large $t_w$ behavior of $R(t,t_w)$
does not reveal that the system remains out of equilibrium in the large time regime.

For what concerns the
exponent $a$, there is a major difference 
with respect to the case of the quench to $T_C$.
Since the FDT in Eq.~(\ref{8.8}) relates only the stationary components $G_{eq}(\tau,T_F)$ and $R_{eq}(\tau,T_F)$,
there is no more any
constraint relating $a$ to $b$, contrary to what happened in the critical quench where 
the equality~(\ref{8.3}) was 
enforced by the multiplicative structure. In other words, in the quench to below $T_C$ the value
of $a$ is decoupled from that of $b$ and, as mentioned in the Introduction, the determination of this exponent
is a difficult and challenging problem, which will be discussed at the end of the chapter. 
In any case, although the actual value of $a$ is, to some extent, a debated issue, 
yet there is general consensus on the statement $a > 0$ for $d > d_L$.
The case of $d=d_L$ stands apart and will be discussed in section \ref{special}.

\subsection{ZFC susceptibility below $T_C$}
\label{ZFC}

It is clear that the additive structure of the response function
generates the analogous form of the ZFC susceptibility
\be
\chi (t,t_w)=\chi_{eq} (\tau,T_F) +\chi_{ag} (t,t_w)
\label{K.1bis}
\ee
where the stationary component 
satisfies, by construction, the equilibrium FDT~(\ref{7.5}).
Inserting the scaling form~(\ref{I.3}) of $R_{ag}$ in the definition of the aging component 
\be
\chi_{ag} (t,t_w) = \int_{t_w}^t dt^{\prime}R_{ag}(t,t^{\prime})
\label{K.2}
\ee
one finds
\be
\chi_{ag} (t,t_w)= t_w^{-a}h_\chi(x,y)
\label{K.3}
\ee
where 
\be
h_\chi(x,y)= x^{-a}{\cal I}(x,y)
\label{K.03}
\ee
and
\be
{\cal I}(x,y) = \int_{1/x}^1 dz\; z^{-(1+a)}h_R (z^{-1},z^{-1}y/x).
\label{K.4}
\ee
Here, we make a first observation which turns out to be quite important in general, when
the data for $\chi_{ag}$ from the simulations are used to measure the exponent $a$~\cite{noiTRM}.
If one seeks to determine $a$ from Eq.~(\ref{K.3}) by looking at the
behavior of $\chi_{ag}$ as $t_w$ is varied and $x$ is kept fixed, one must
be aware that the $t_w$ dependence coming from $y=t_0/t_w$ may play a role. 
In other words, $t_0$ may act as a dangerous irrelevant variable through a
mechanism quite similar to the one causing the breakdown of hyperscaling in static
critical phenomena above the upper critical dimensionality $d_U$~\cite{dangerous}. 
In those cases where
analytical calculations can be carried out with arbitrary $d$~\cite{EPJ,Berthier,Ninfty}
there exists a value $d^*$ of the dimensionality such that
the limit for $y \rightarrow 0$ of the integral ${\cal I}(x,y)$ is
finite for $d < d^*$, while for $d>d^*$ there is a singularity of the type
\be
{\cal I}(x,y) = (y/x)^{-c}\;\widehat{{\cal I}}(x)
\label{dang}
\ee
with $c>0$, which becomes logarithmic $(c=0)$ for $d=d^*$. Hence, $d^*$ plays the same role as 
$d_U$ in critical phenomena, but it is clearly unrelated to $d_U$. 
The assumption is that this is a generic feature of the relaxation to below $T_C$.
Then, Eq.~(\ref{K.3}) can be rewritten as 

\be
\chi_{ag} (t,t_w)= t_w^{-a_\chi}\widehat{h}_\chi(x)
\label{K.5}
\ee
with
\be
a_\chi= \left \{ \begin{array}{ll}
        a, \qquad $for$ \qquad d<d^* \\
        a, \qquad $with log corrections for$ \qquad d=d^* \\
        a-c, \qquad $for$ \qquad d>d^* 
        \end{array}
        \right .
        \label{K.50}
        \ee
and for large $x$ 
\be
\widehat{h}_\chi(x)\sim x^{-a_\chi}
\label{K.6}
\ee
which, in turn, implies 
\be
\chi_{ag} (t,t_w) \sim t^{-a_\chi}
\label{K.7}
\ee
for large $t$.
The analogy with the breaking of hyperscaling is quite close since, as we shall see,
there is a dependence on $d$ of $a_\chi$ for $d<d^*$, which disappears
for  $d>d^*$.

The second observation concerns the value of $a_\chi$ and the asymptotic behavior
of $\chi(t,t_w)$. From Eq.~(\ref{K.1bis}) follows
\be
\lim_{t \to \infty} \chi(t,t_w) = \lim_{\tau \to \infty}  \chi_{eq}(\tau,T_F) + \lim_{t \to \infty} \chi_{ag}(t,t_w)
\label{K.71}
\ee
and, recalling Eqs.~(\ref{3.024}) and~(\ref{K.7}), this gives
\be
\lim_{t \to \infty} \chi(t,t_w) = \chi_{st}(T_F) + \chi^*
\label{K.72}
\ee
with
\be
    \chi^* = \left \{ \begin{array}{ll}
        0 \qquad $for$ \qquad a_\chi > 0 \\
        \lim_{x \to \infty}\widehat{h}_\chi(x) \qquad $for$ \qquad a_\chi = 0. 
        \end{array}
        \right .
        \label{K.11}
        \ee
Hence, the ZFC susceptibility reaches the equilibrium value 
for  $a_\chi>0$, but not for  $a_\chi=0$.
In other words, for $a_\chi > 0$ the contribution of the slow degrees of freedom disappears asymptotically,
while for $a_\chi =0$ there remains an extra contribution $\chi^*$ on top of the equilibrium one.
This is a very interesting phenomenon. Rewriting Eq.~(\ref{2.281}) as $T_F\chi_{st}(T_F) = C_{eq}(0,T_F) -M^2$
and recalling that $M^2$ plays the role of the Edwards-Anderson order parameter,
there is a formal similarity with what happens in the mean-field theory of spin 
glasses, where the large time limit of the ZFC susceptibility
can be written~\cite{Hertz} exactly as in Eq.~(\ref{K.72}). The substantial difference 
is that in the spin glass
case $\chi^*$ is an equilibrium quantity, whose appearence is due to replica simmetry breaking~\cite{MPV}.
As a matter of fact, the observation of $\chi^* >0$ in the simulations of finite dimensional spin glasses is
taken as evidence of replica simmetry breaking~\cite{PRR}. Here, instead, $\chi^*$ 
is the difference between the large time limit of the ZFC susceptibility and the same quantity
computed from equilibrium statistical mechanics. It is a quantity of purely
dynamical origin which appears, as we shall see, in the quench to $(d_L,T_F=0)$ revealing the 
strong out of equilibrium nature of the relaxation.

\subsection{Fluctuation-dissipation ratio (FDR)}

If the system is not in equilibrium the FDT does not hold and the violation of the theorem
can be used as a measure of the deviation from equilibrium. This idea was 
implemented by Cugliandolo and Kurchan~\cite{CK} through the introduction of the FDR
\be
X(t,t_w)= {T_F R(t,t_w) \over \partial_{t_w} C(t,t_w)}
\label{9.1}
\ee
which satisfies $X(t,t_w) \equiv 1$ in equilibrium and $X(t,t_w) \neq 1$ off-equilibrium.
Formally, the FDR allows to define the temperature-like quantity
\be
T_{eff}(t,t_w)= {T_F \over X(t,t_w)}
\label{9.1bis}
\ee
whose interpretation as an effective temperature, however, requires some care~\cite{Peliti}.

As we have seen, the characterization of systems which remain out of equilibrium for
arbitrary long times requires the exploration of the various asymptotic regimes reached
as $t_w \rightarrow \infty$. An efficient way of doing this is through the
reparametrization of the time $t$ in terms of the autocorrelation function.
For fixed $t_w$, $C(t,t_w)$ is a monotonously decreasing function of $t$. Hence, inverting
with respect to $t$, the function
$\widehat{X}(C,t_w)= X(t(C,t_w),t_w)$ is obtained, whose limit for fixed $C$
\be
\lim_{t_w \to \infty} \widehat{X}(C,t_w) = {\cal X}(C)
\label{9.2}
\ee
defines the limit FDR in the time sector characterized by the chosen value of $C$ and
the associated effective temperature is given by
\be
{\cal T}(C) = {T_F \over {\cal X}(C)}.
\label{9.2bis}
\ee
The correspondence between values of $C$ and the short and long
time regimes will be clarified below.

Inserting $X(t,t_w)$ into the definition~(\ref{3.023}) of the ZFC susceptibility
\be
T_F \chi(t,t_w) = \int_{t_w}^t dt^{\prime} \; \widehat{X}(C(t,t^{\prime}),t^{\prime})
{\partial \over \partial t^{\prime}} C(t,t^{\prime})
\label{9.3}
\ee
and for values of $t_w$ so large that Eq.~(\ref{9.2}) can be used under the integral,
the parametric representation of $\chi$ is obtained 
\be
T_F\widehat{\chi}(C(t,t_w)) = \int_{C(t,t_w)}^{C(t,t)} dC^{\prime}  \; {\cal X}(C^{\prime}).
\label{9.4}
\ee
Differentiating with respect to $C$ this gives
\be
-T_F{d \widehat{\chi}(C) \over dC} ={\cal X}(C) 
\label{9.5}
\ee
which relates the limit FDR to the slope of $\widehat{\chi}(C)$.
This is the most commonly used way of estimating the FDR, due to the relative ease of computing
$\widehat{\chi}(C)$ numerically.
Notice that the integrated form~(\ref{7.5}) of the FDT 
is recovered from Eq.~(\ref{9.4}) when ${\cal X}(C) \equiv 1$. In that case the plot of 
$T_F\widehat{\chi}(C)$ is a straight line with slope $-1$, the so called trivial plot, which
is the hallmark of equilibrium. Off-equilibrium behavior is conveniently detected through the
deviations from the trivial plot.

\subsection{Parametric plots}

\begin{figure}[h]
\begin{center}
\includegraphics*[scale = 0.5]{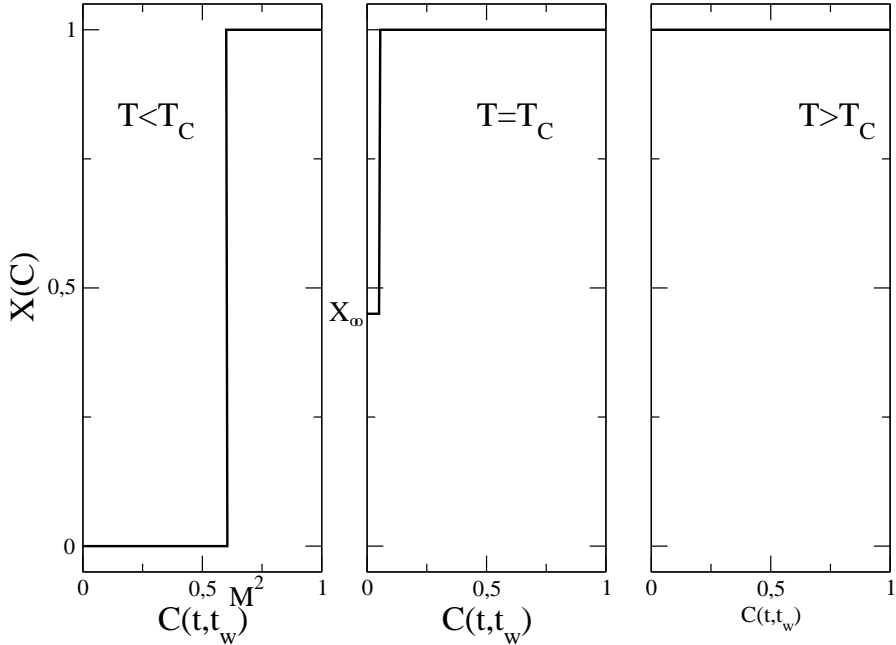}
\end{center}
\caption{\label{fig10} The limit FDR ${\cal X}(C)$.}
\end{figure}

The shape of the parametric plots  (Fig.\ref{fig10}) can be derived from general considerations.
In the quench to above $T_C$ the system equilibrates in a finite time and it is straightforward to obtain
\be
{\cal X}(C) \equiv 1
\label{10.00}
\ee
since $X(t,t_w) \equiv 1$ for $t_w > t_{eq}$.
In the case of the critical quench the outcome is almost the same, but the derivation
is less straightforward. From Eqs.~(\ref{5.140},\ref{8.3},\ref{8.100}) one finds
\be
X(t,t_w)= F(x,y)
\label{10.1}
\ee
where
\be
F(x,y)= -T_C{f_R(x,y) \over f_{\partial C}(x,y)}
\label{10.01}
\ee
and 
\be
f_{\partial C}(x,y)= b_c f_C(x,y) + \left [ x {\partial \over \partial x}f_C(x,y) +
y {\partial \over \partial y}f_C(x,y) \right ].
\label{10.2}
\ee
Inverting with respect to $x$ the form~(\ref{5.190}) of the autocorrelation function  
\be
x(C)= \left \{ \begin{array}{ll}
        \infty, \qquad $for$ \qquad C=0 \\
        1,  \qquad $for$ \qquad 0 < C \leq C_{eq}(0,T_C)
        \end{array}
        \right .
        \label{10.02}
        \ee
and inserting into Eq.~(\ref{10.1}), the limit FDR is obtained 
\be
{\cal X}(C) = \left \{ \begin{array}{ll}
        {\cal X}_{\infty}, \qquad $for$ \qquad C=0 \\
        1,  \qquad $for$ \qquad 0 < C \leq C_{eq}(0,T_C)
        \end{array}
        \right .
        \label{10.03}
        \ee
where
\be
{\cal X}_{\infty} = \lim_{y \to 0}\lim_{x \to \infty} F(x,y)= {T_C A_R \over A_C (\lambda/z_c - b_c)}
\label{10.04}
\ee
is a new universal quantity characteristic of the critical relaxation~\cite{Godreche02,Godreche00}.
The second line of~(\ref{10.03}) comes from $F(1,0)=T_Cg_R(1)/b_cg_C(1)$, together with the FDT
requirement~(\ref{8.6}). The above result illustrates quite well the usefulness of the FDR and
of the parametric plot in the precise characterization of the off-equilibrium relaxation. Eq.~(\ref{10.02})
shows that in the critical quench all values of $C>0$ correspond to the short time regime, while the large time 
corresponds to $C=0$ and in the latter regime the system remains off-equilibrium since, in general,
${\cal X}_{\infty} < 1$ (center panel of Fig.\ref{fig10}).

In the quench to below $T_C$, from the additive structures of $C$ and $R$ follows
\begin{eqnarray}
X(t,t_w) & = & T_F\left [ {R_{eq}(\tau,T_F) \over \partial_{t_w} G_{eq}(\tau,T_F) 
+ \partial_{t_w} C_{ag}(t,t_w)} \right. \\ \nonumber
& + & \left. {R_{ag}(t,t_w) \over \partial_{t_w} G_{eq}(\tau,T_F) + \partial_{t_w} C_{ag}(t,t_w)}\right ].
\label{10.3}
\end{eqnarray}
Recalling that $R_{eq}$ and $R_{ag}$ are not simultaneously different from zero when $t_w$ is large,
the first term in the brackets contributes in the short time regime and the second one in the large time
regime, yielding
\be
    X(t,t_w)= \left \{ \begin{array}{ll}
        1 \qquad $for$ \qquad x=1 \\
        t_w^{-a}H(x,y) \qquad $for$ \qquad x>1 
        \end{array}
        \right .
        \label{10.4}
        \ee
where 
\be
H(x,y)= -T_F{h_R(x,y) \over h_{\partial C}(x,y)}
\label{10.4bis}
\ee
with
\be
h_{\partial C}(x,y)= \left [ x {\partial \over \partial x}h_C(x,y) +
y {\partial \over \partial y}h_C(x,y) \right ].
\label{10.5}
\ee
Again, inverting with respect to $x$ the autocorrelation function~(\ref{5.22}) 
\be
x(C)= \left \{ \begin{array}{ll}
        h_C^{-1}(C), \qquad $for$ \qquad C < M^2 \\
        1,  \qquad $for$ \qquad M^2 \leq C \leq C_{eq}(0,T_F)
        \end{array}
        \right .
        \label{10.6}
        \ee
and inserting into~(\ref{10.4}) one finds
\be
         X(C,t_w)= \left \{ \begin{array}{ll}
         t_w^{-a}H(h_C^{-1}(C),y), \qquad $for$ \qquad C < M^2 \\
         1, \qquad $for$ \qquad M^2 \leq C \leq C_{eq}(0,T_F) 
        \end{array}
        \right .
        \label{10.7}
        \ee
whose $t_w \rightarrow \infty$ limit for $d > d_L$, that is when $a>0$, gives
\be
    {\cal X}(C)= \left \{ \begin{array}{ll}
         0 \qquad $for$ \qquad C < M^2 \\
         1 \qquad $for$ \qquad M^2 \leq C \leq C_{eq}(0,T_F). 
        \end{array}
        \right .
        \label{10.8}
        \ee
Here, the off-equilibrium character of the relaxation is expanded and enhanced, 
with respect to the critical quench (left panel of Fig.\ref{fig10}). 
The quasi equilibrium in the short time regime is limited to values of $C$ above
the Edwards-Anderson plateau and the deviation from equilibrium for large time is more pronounced, 
since the FDR vanishes.

\begin{figure}[h]
\begin{center}
\includegraphics*[scale = 0.5]{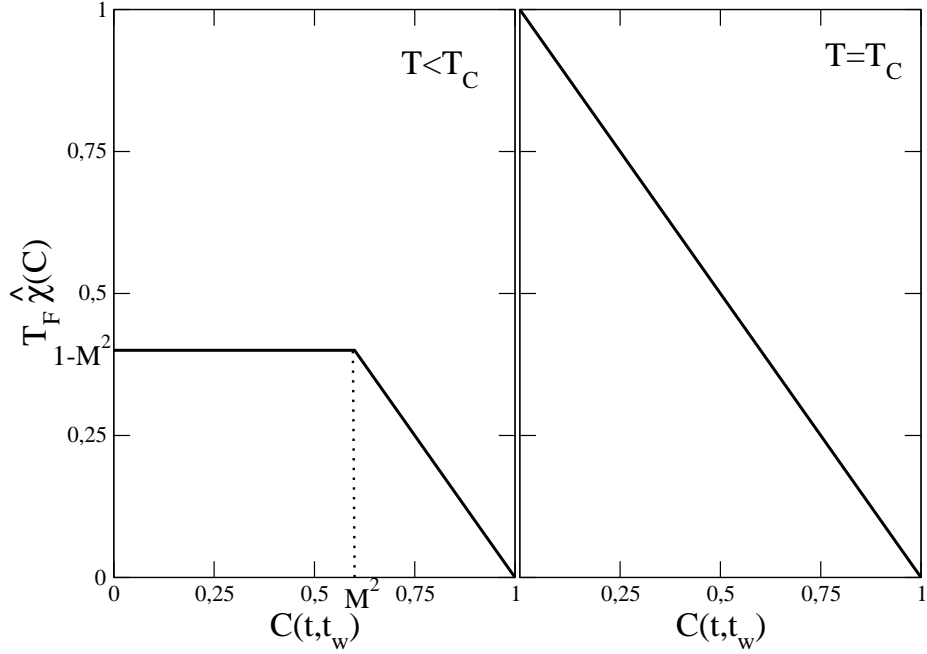}
\end{center}
\caption{\label{fig11} Parametric plots of the ZFC susceptibility $\hat{\chi}(C)$ with $C_{eq}(0,T_F)=1$}
\end{figure}

The corresponding parametric representations of the ZFC susceptibility are easily obtained (Fig.\ref{fig11}),
by integration.
In the quenches to above and to $T_C$, both Eqs.~(\ref{10.00}) and~(\ref{10.03}) yield the 
same trivial plot
\be
T_C \hat{\chi}(C) = C_{eq}(0,T_C) - C.
\label{10.9}
\ee
Notice that if one goes back differentiating with respect to $C$, one finds identically
${\cal X}(C)=1$ for all values of $C$, above and at $T_C$. This clarifies that in
order to uncover the existence of the non trivial FDR ${\cal X}_\infty$, the order
of the limits in Eq.~(\ref{10.04}) is crucial.
In the quench to below $T_C$, instead, the departure from the trivial behavior is most
evident
\be
    T_F\hat{\chi}(C)= \left \{ \begin{array}{ll}
         C_{eq}(0,T_F)-M^2\qquad $for$ \qquad C < M^2 \\
         C_{eq}(0,T_F) - C \qquad $for$ \qquad M^2 \leq C \leq C_{eq}(0,T_F) 
        \end{array}
        \right .
        \label{10.10}
        \ee
since the plot is flat for $C < M^2$. 
This plot shows at glance that the susceptibility
equilibrates, while the autocorrelation function does not.
The rise from zero to $C_{eq}(0,T_F)-M^2$ in the left panel
of Fig.\ref{fig11} shows the saturation of $\widehat{\chi}(C)$ to the static 
value $T_F\chi_{st}(T_F)=G_{eq}(0,T_F)$ as $C$ decays to the Edwards-Anderson plateau.
For larger times the susceptibility remains fixed at the equilibrium value in the flat portion of
the plot, while $C$ falls below the plateau, according to the weak ergodicity
breaking scenario.

The results above described are universal, since all the 
non universal features of $F(x,y)$ and $H(x,y)$ have been eliminated in the $t_w \rightarrow \infty$
limit. Therefore, in all the quenches to $T_C$ the parametric plots of ${\cal X}(C)$ and
$T_C\hat{\chi}(C)$ are trivial, except for the value of ${\cal X}_{\infty}$, and in all
quenches to below $T_C$ the deviation from the trivial plot takes the form
of the flat behavior below the Edwards-Anderson plateau.
For the parametric plots in the quench to $(d_L,T_F)$ it is not possible to make 
statements of such generality. Comments will be made in the next section.

Finally, few words about the effective temperature. From Eqs.~(\ref{9.2bis}) and~(\ref{10.00})
follows that ${\cal T}(C)$ coincides with the temperature of the thermal bath $T_F$,   
when the plot of  $T_F\hat{\chi}(C)$ is trivial. For $T_F > T_C$ this happens for all values of
$C$. For  $T_F = T_C$ this happens for all values of $C$, except $C=0$, where
${\cal T}(0) = T_F/{\cal X}_{\infty} > T_F$ and, finally, for $T_F < T_C$ 
\be
    {\cal T}(C)= \left \{ \begin{array}{ll}
         \infty \qquad $for$ \qquad C < M^2 \\
         T_F \qquad $for$ \qquad M^2 \leq C \leq C_{eq}(0,T_F). 
        \end{array}
        \right .
        \label{10.11}
        \ee
The latter result is suggestive that in the quench
to below $T_C$, while the fast degrees of freedom thermalize, the
slow ones do not interact at all with the thermal bath and keep on remaining to the temperature $T_I$ 
of the initial condition.

\subsection{A special case: the quench to $(d_L,T_F=0)$}
\label{special}

An interesting and non trivial situation arises if the system is at the 
lower critical dimensionality $d_L$
and the quench is made at $T_F=0$. As anticipated in the Introduction,
this process can be regarded as the limit for $d \rightarrow d_L$ either of 
critical quenches or of quenches made to below $T_C$. 
In the first case from Eq.~(\ref{10.03})
\be
{\cal X}(C) = \left \{ \begin{array}{ll}
        0, \qquad $for$ \qquad C=0 \\
        1,  \qquad $for$ \qquad 0 < C \leq C_{eq}(0,T_C)
        \end{array}
        \right .
        \label{11.1}
        \ee
since ${\cal X}_{\infty}=0$, while in the second case from Eq.~(\ref{10.8})
\be
    {\cal X}(C)= \left \{ \begin{array}{ll}
         0 \qquad $for$ \qquad C < C_{eq}(0,T_F)\\
         1 \qquad $for$ \qquad C =C_{eq}(0,T_F)
        \end{array}
        \right .
        \label{11.2x}
        \ee
since $C_{eq}(0,T_F)=M^2$ at $T_F=0$. Hence, two very different results are obtained
and there is the problem of which is the correct one. What can be stated on general
grounds is that the quench to $(d_L,T_F=0)$ is akin to a quenche to below $T_C$,
since $G_{eq}(\tau,T_F) \equiv 0$ at $T_F=0$ and, therefore, $C(t,t_w)$ must necessarily have
the additive structure, otherwise would vanish identically. Hence, Eq.~(\ref{11.1}) 
can be discarded.
However, from the analytical~\cite{EPJ,Lippiello,Godreche1d,Ninfty}
and numerical~\cite{universality,generic}
evidence accumulated so far, ${\cal X}(C)$ does not obey Eq.~(\ref{11.2x}) either.
Rather, ${\cal X}(C)$ appears to be a non-trivial and non-universal smooth function.  
Such a behavior is compatible with Eq.~(\ref{10.7}) if $a=0$, which yields the smooth behavior
\be
{\cal X}(C) = H(h_C^{-1}(C),0) 
\label{11.3}
\ee
preserving all the non-universal features of
$h_C(x,0)$ and $H(x,0)$. Therefore, in the quench to $(d_L,T_F=0)$, although belonging 
to the class of the quenches to below $T_C$, there are peculiarities induced by the vanishing 
of the exponent $a$.

\section{Models}

The general concepts introduced in the previous sections will now be
illustrated through analytical and numerical results for specific models,
limiting the discussion to the case of non conserved order parameter.
The models considered are 

\begin{itemize}

\item the Ising model with the Hamiltonian 
\be
{\cal H}[s_i] = J \sum_{<ij>}s_i s_j
\label{2.1}
\ee
where the sum runs over the pairs $<ij>$ of the nearest neighbors
spins with the ferromagnetic coupling $J <0$. 
The time evolution with Glauber single spin flip dynamics~\cite{Glauber} is governed by
the master equation 
\be
{\partial \over \partial t} P([s],t) = \sum_i \left \{ w(-s_i)P([R_is],t) - w(s_i)P([s],t) \right \}
\label{3.1}
\ee
where $P([s],t)$ is the probability of realization of the spin configuration $[s]$ at the time $t$ and
$[R_is]$ is the configuration with the $i$-th spin reversed,
\be
w(s_i)= {1 \over 2\tau_0}[1-s_i \tanh (E_i/T)]
\label{3.2}
\ee
is the transition rate from $[s]$ to $[R_is]$, $\tau_0$ is a constant and
$E_i=h_i + J\sum_{j \in <i>}s_j$ is the sum of the external and the local field on the spin $s_i$,
due to its nearest neighbors.

\item the continous Ginzburg-Landau-Wilson (GLW) model with the Hamiltonian
\be
{\cal H}[\vec \phi (\vec x)] = \int_V d \vec x \;
\left [ {1 \over 2} (\nabla \vec \phi )^2 + {\mu \over 2} \vec \phi^2 
+ {u \over 4 N} (\vec \phi^2 )^2 \right ]
\label{2.2}
\ee 
where $\vec \phi(\vec x)= (\phi_1 (\vec x),..,\phi_N (\vec x))$ is an $N$-component
vector order parameter,
the integral is taken over the volume $V$ and $\mu <0$, $u >0$ are constants.
The purely relaxational dynamics (model A in the classification
of Hohenberg-Halperin~\cite{HH}) is governed by the Langevin equation~\cite{Goldenfeld,Chaikin}
\be
{\partial \vec \phi(\vec x,t)\over \partial t} = -{\delta {\cal H}[ \vec \phi]
\over \delta \vec \phi(\vec x,t)} + \vec \eta(\vec x,t)
\label{3.4L}
\ee
where $\vec \eta(\vec x,t)$ is a Gaussian white noise with expectations
\be
\langle \vec \eta(\vec x,t) \rangle = 0
\label{3.5}
\ee
and
\be
\langle \eta_{\alpha}(\vec x,t)  \eta_{\beta}(\vec x^{\prime},t^{\prime})\rangle =
2T\delta_{\alpha,\beta}\delta(x-x^{\prime})\delta(t-t^{\prime}). 
\label{3.6}
\ee

\end{itemize}

\noindent These models can be solved exactly only in a very limited number of cases:
in $d=1$ for the Ising model and for arbitrary $d$ in the vector GLW model, after
taking the $ N \rightarrow \infty$ (large $N$)~\cite{Chaikin,Dean,Godreche00,Ninfty,Yoshino2} limit. 
Otherwise, one must resort either to
numerical simulations or to approximation methods, as discussed in the Introduction.

\subsection{Large $N$ model}
\label{largeN}

When the number $N$ of order parameter components goes to infinity, the mean-field-like 
linearization of the GLW Hamiltonian, obtained by the replacement
$(\vec \phi^2)^2 \Rightarrow  2\langle\vec \phi^2 \rangle \vec \phi^2$ in Eq.~(\ref{2.2}),
becomes exact, both for statics and dynamics, with the proviso that the average
$\langle\vec \phi^2 \rangle$ must be computed self-consistently.

Therefore, in the large $N$ limit the equation of motion
for the Fourier transform of the order parameter 
$ \vec \phi (\vec k) = \int _V d^d x \vec \phi(\vec x) 
\exp (i \vec k \cdot \vec x)$ takes the linear form
\be
\frac{\partial \vec \phi (\vec k ,t) }{\partial t} 
= - [k^2 + I(t) ] \vec \phi( \vec k,t) + \vec \eta (\vec k,t)
\label{02.5}
\ee
where 
\be
I(t) = \mu + \frac{u}{N} \langle \vec \phi ^2(\vec x,t)\rangle 
\label{02.7}
\ee
and the average is taken over the noise
\be
\left \{ \begin{array}{ll}
      \langle \vec \eta (\vec k, t) \rangle = 0  \\ 
      \langle \eta _\alpha (\vec k,t) \eta _\beta (\vec k',t') \rangle = 2T_F
      \delta _{\alpha,\beta}V \delta_{\vec k +\vec k',0} \delta(t-t') 
                       \end{array}                                        
               \right . 
\label{02.6}
\ee
and the initial condition
\be
\left \{ \begin{array}{ll}
      \langle \vec \phi (\vec k) \rangle_I = 0  \\ 
      \langle \phi _\alpha (\vec k) \phi _\beta (\vec k^{\prime}) \rangle_I = 
      \Delta \delta _{\alpha,\beta} V \delta_{\vec k +\vec k^{\prime},0}.
                       \end{array}                                        
               \right . 
\label{02.3}
\ee

\subsubsection{Statics}

If the volume $V$ is kept finite the system equilibrates in a 
finite time $t_{eq}$ and the order parameter probability distribution
 reaches the Gibbs state 
\be
P_{G} [\vec \phi(\vec k)] = \frac {1}{Z} e^{ -\frac {1}{2 T_F V} 
\sum _{\vec k} ( k^2+ \xi_F ^{-2}) \vec \phi(\vec k) \cdot \vec \phi(-\vec k)  }
\label{02.8}
\ee
where $ \xi_F $ is the correlation length defined by 
the static self-consistency condition 
\be
\xi_F ^{-2} = \mu+\frac{u}{N} \langle \vec \phi ^2(\vec x) \rangle _{G}.
\label{02.9}
\ee

In order to analyze the properties of
$P_{G} [\vec \phi(\vec k)]$ it is necessary to extract 
the dependence of $\xi_F$ on $T_F$ and $V$. Evaluating the average,
the above equation takes the form
\be
\xi_F ^{-2} = \mu+\frac{u}{V} \sum _{\vec k} \frac{T_F}{k^2+\xi_F ^{-2}} 
\label{02.10}
\ee
whose solution is well known~\cite{Baxter,Ninfty}. There exists the critical temperature 
\be
T_C= -{\mu \over u}(4 \pi)^{d/2} \Lambda^{2-d} (d-2)/2
\label{T_C}
\ee
where $\Lambda$ is an high momentum cutoff.
For $T_F>T_C$ the solution of Eq.~(\ref{02.10}) is independent of the volume,
while for $T_F \leq T_C$ depends on the volume
\be
    \xi_F \left \{ \begin{array}{ll}
         \sim \large ( \frac{T_F-T_C}{T_C} \large )^{-\nu} \qquad $for$ \qquad T_F > T_C\\
         \sim V^{1/d} \qquad $for$ \qquad T_F = T_C\\ 
         = \sqrt{\frac {M^2 V}{T_F}} \qquad $for$ \qquad  T_F < T_C
        \end{array}
        \right .
        \label{11.2}
        \ee
where 
\be 
M^2= M_0^2 \left ( \frac {T_C - T_F}{T_C} \right )
\label{M2}
\ee
is the square of the spontaneous
magnetization, $M_0^2 = -\mu/u$ and $\nu = 1/(d-2)$. Notice that from Eq.~(\ref{T_C})
follows that the critical line $T_C(d)$ in Fig.\ref{fig1} is
a straight line and that $d_L=2$.

Let us now see what are the implications for the equilibrium state. 
As Eq.~(\ref{02.8}) shows, the individual Fourier components are 
independent random variables, gaussianly distributed with zero 
average for all temperatures. The variance of each mode is given by 
\be
\frac{1}{N} \langle \vec \phi(\vec k) \cdot \vec \phi(-\vec k) \rangle _{G} = 
V C_{eq}(\vec k)
\label{02.18}
\ee
where 
\be
C_{eq}(\vec k) = \frac{ T_F}{k^2+ \xi_F^{-2}}
\label{02.19}
\ee
is the equilibrium structure factor. For $T_F >T_C$, 
all $\vec k$ modes behave in the same way, 
with the variance growing linearly with the volume. For $T_F \leq T_C$, 
instead, $\xi_F^{-2}$ is negligible
with respect to $k^2$ except at $\vec k=0$, yielding  
\bea
C_{eq}(\vec k)  =   \left \{   
                       \begin{array}{ll}
                          \frac{ T_C}{k^2} (1 -\delta _{\vec k,0}) + 
                          \kappa V^{2/d} \delta _{\vec k,0}      & 
                                 \mbox{, for $T_F=T_c$ }\\ 
        \frac{T_F}{k^2 }(1-\delta _{\vec k,0})+ M^2 V \delta _{\vec k,0} &
				 \mbox{, for $T_F<T_c$} 
                       \end{array}
		    \right .
\label{02.20}
\eea
where $\kappa$ is a constant. Therefore, for $T_F \leq T_C$ the
$\vec k=0$ mode behaves differently from all the other modes,
since the variance grows faster than linear with the volume.
In particular, for $T_F<T_C$ the Gibbs state 
takes the form 
\be
P_{G}[\vec \phi (\vec k) ] = \frac {1} {Z} 
e ^{-\frac{\vec \phi ^2(0)}{2 M^2 V^2}}  
e^{ -\frac{1}{2T_FV} \sum_{\vec k} k^2 \vec \phi (\vec k) \cdot
\vec \phi (-\vec k) }
\quad .
\label{02.21}
\ee
Therefore, crossing $T_C$ there is a transition from the usual disordered
high temperature phase to a low temperature phase which, instead of being
the mixture of broken symmetry states, is characterized by a 
macroscopic variance in the Gaussian distribution of the $\vec k=0$ mode.
In place of the transition from disorder to order, it is more appropriate
to speak of the condensation of fluctuations in the $\vec k=0$ mode.
The distinction between the condensed phase and the mixture of pure states,
has been discussed in detail in Ref. \cite{Castellano97}.

Despite the difference in the mechanism of the transition,
from Eqs.~(\ref{02.19}) and~(\ref{02.20}) it is easy to see that     
the correlation function follows the same pattern outlined in
general in Eqs.~(\ref{2.9},\ref{2.14},\ref{2.15}),
with $\eta=0$ and $M^2$ given by~(\ref{M2}). Furthermore, the splitting~(\ref{2.20}) of the order parameter
\be
\vec \phi (\vec x) = \vec \psi (\vec x) + \vec \sigma
\label{02.22}
\ee
now can be carried out explicitely taking
\be
\vec \sigma = \frac{1}{V}\vec \phi (\vec k=0)
\label{02.23}
\ee
and 
\be 
\vec \psi (\vec x) = \frac{1}{V} \sum _{\vec k \neq 0 } 
\vec \phi (\vec k) e^{i\vec k \cdot \vec x}.
\label{02.24}
\ee
Then, rewriting the Gibbs state as
\be
P_{G}[\vec \phi (\vec x) ] = P(\vec \sigma) P[\vec \psi (\vec x) ] 
\label{02.24.1}
\ee
with  
\be
P(\vec \sigma) = \frac{1}{(2 \pi M^2)^{N/2}} e^{- \frac{\vec \sigma ^2  }
{2 M^2}} \quad 
\label{02.25}
\ee
and
\be
P[\vec \psi (\vec x)] = \frac{1}{Z} e^{-\frac{1}{2T_F}
\int_V d^d x (\nabla \vec \psi )^2}
\label{02.25.1}
\ee
the two contributions in Eq.~(\ref{2.14}) are given by 
\be
G_{eq}(\vec x - \vec x' ,T_F) =  \frac{1}{N} 
\langle \vec \psi (\vec x) \cdot \vec \psi (\vec x') \rangle _{G} 
\label{02.27}
\ee
and
\be
M^2 =  \frac{1}{N} 
\langle \vec \sigma \cdot \vec \sigma \rangle _{G}. 
\label{02.27bis}
\ee

\subsubsection{Dynamics} \label{sec3}

Taking advantage of the rotational symmetry and of the effective decoupling of the vector
components, from now on we shall drop vectors
and refer to the generic order parameter component.
The formal solution of the equation of motion~(\ref{02.5}) reads
\be
\phi (\vec k,t) = R(\vec k,t,0)\phi _0(\vec k)+
\int _0 ^t dt' R(\vec k,t,t')\eta (\vec k,t')
\label{03.5}
\ee
where
\be
R(\vec k,t,t')=\frac{Y(t')}{Y(t)}e^{-k^2(t-t')}
\label{03.6}
\ee
is the response function, $\phi _0(\vec k)=\phi (\vec k,0)$ is the initial value of the order parameter and
\be
Y(t)=\exp \{\int _0 ^t ds I(s)\}
\label{03.6.1}
\ee
is the key quantity in the exact solution of the model. 
In order to find it, notice that from the definition of $Y(t)$ follows
\be
\frac{dY^2(t)}{dt}=2\left [ \mu+u\langle \phi ^2(\vec x,t)\rangle \right ]
Y^2(t).
\label{03.9}
\ee
Writing $\langle \phi ^2(\vec x,t)\rangle $ in terms of the 
structure factor
\be
\langle \phi ^2(\vec x,t)\rangle =\int \frac {d^dk}{(2\pi )^d}
C(\vec k,t)e^{-\frac{k^2}{\Lambda ^2}}
\label{03.10}
\ee
and using~(\ref{03.5}) to evaluate $C(\vec k,t)$  
\be
C(\vec k,t)=R^2(k,t,0)\Delta+2T_F\int _0 ^t dt' R^2 (\vec k,t,t')
\label{03.11}
\ee
from~(\ref{03.9}) one obtains the integro-differential 
equation 
\be
\frac{dY^2(t)}{dt}=2\mu Y^2(t)+2u\Delta J\left (t+\frac{1}{2\Lambda ^2}\right )
+4uT_F\int _0 ^t dt' J\left (t-t'+\frac{1}{2\Lambda ^2}\right ) Y^2(t')
\label{03.12}
\ee
where
\be
J(x)\equiv \int \frac{d^dk}{(2\pi )^d}e^{-2k^2x}=(8\pi x)^{-\frac{d}{2}}.
\label{03.13}
\ee
Solving~(\ref{03.12}) by Laplace transform \cite{Newman90,Godreche00}, 
the leading behavior of $Y(t)$ for large time is given by~\cite{Ninfty,Godreche00}
\be
Y(t)=\left \{ \begin{array}{ll}
                              A_ae^{t/\xi_F}    & \mbox{, for $T_F>T_C$} \\
                              A_ct^{(d-4)/4} & \mbox{, for $T_F=T_C$} \\
			      A_bt^{-\frac{d}{4}}  & \mbox{, for $T_F<T_C$}
                              \end{array}
       \right .
\label{03.14}
\ee
where $A_a, A_b, A_c$ are constants.

\subsubsection{Splitting of the order parameter}

The solution of the model will now be used to show 
how the splitting~(\ref{6.4}) of the order parameter into the sum
of two independent contributions, with the properties~(\ref{6.5}) and~(\ref{6.6}), can be explicitely carried out
and, at the same time, to give a derivation of the properties of $C(t,t_w)$ and $R(t,t_w)$ which have been 
stated in general in the previous sections.
From the multiplicative property of the response function
\be
R(\vec k,t,t') R(\vec k,t',t^*)= R(\vec k,t,t^*)
\label{04.4}
\ee
for any ordered triplet of times $t^*<t'<t$, it is easy to show that the solution~(\ref{03.5})
can be rewritten as the sum of two statistically
independent contributions
$\phi (\vec k,t) =\psi (\vec k,t)+\sigma(\vec k,t)$
with
\be
\sigma(\vec k,t)= R(\vec k,t,t^*)\phi (\vec k,t^*)
\label{04.6}
\ee
and
\be
\psi (\vec k,t)=\int _{t^*}^t dt' R(\vec k,t,t')\eta (\vec k,t')
\label{04.7}
\ee
since for $0\leq t^*<t$, $\phi (\vec k,t^*)$ and $\eta (\vec k,t)$ are
independent by causality. In other words, the order parameter at the
time $t$ is split into the sum of a component $\sigma (\vec k,t)$,
driven by the fluctuations of the order parameter at the earlier time
$t^*$, and a component $\psi (\vec k,t)$, driven by the thermal history
between $t^*$ and $t$.
Recall that $t^*$ can be chosen arbitrarily between the initial 
time of the quench ($t=0$) and the observation time $t$. With the
particular choice $t^*=0$, the component $\sigma (\vec k,t)$ is
driven by the fluctuations in the initial condition~(\ref{02.3}). 
The $\psi $ component
describes fluctuations of thermal origin while the $\sigma $ component,
as it will be clear below, if $t^*$ is chosen sufficiently large 
describes the local
condensation of the order parameter.

From~(\ref{02.6}) and~(\ref{02.3}) follows
$\langle \sigma (\vec k,t)\rangle=\langle \psi (\vec k,t)\rangle=0$,
while the two time structure factor separates into the sum
\be
C(\vec k,t,t_w)=C_\sigma (\vec k,t,t_w,t^*)+C_\psi (\vec k,t,t_w,t^*)
\label{04.11}
\ee
with 
\be
C_\sigma (\vec k,t,t_w,t^*)=
R(\vec k,t,t^*) R(\vec k,t_w,t^*)C(\vec k,t^*)
\label{04.12}
\ee
and
\be
C_\psi (\vec k,t,t_w,t^*)
=2T_F\int _{t^*}^{t_w}dt'' R(\vec k,t,t'')R(\vec k,t_w,t'').   
\label{04.13}
\ee
The $t^*$ dependence of the two contributions, of course, cancels out in the sum.
Then, going to real space, the autocorrelation function
can be rewritten as
\be
C(t,t_w)=C_\sigma (t,t_w,t^*)+C_\psi (t,t_w,t^*)
\label{04.14}
\ee
with
\be
C_\psi (t,t_w,t^*)=\frac{2T_F}{Y(t)Y(t_w)}\int _{t^*}^{t_w} dt''
J\left ( \frac{t+t_w}{2}-t''
+\frac{1}{2\Lambda ^2}\right ) Y^2(t'')
\label{04.18}
\ee
and
\be
C_\sigma (t,t_w,t^*)=\frac{Y^2(t^*)}{Y(t)Y(t_w)}\int \frac{d^dk}{(2\pi )^d}
e^{-k^2\left (t+t_w-2t^*+\frac{1}{\Lambda ^2}\right ) }C(\vec k,t^*).
\label{04.15}
\ee
Assuming that $t$ and $t_w$ are sufficiently larger than $t^*$, and {\it a fortiori}
of the microscopic time $t_0=\Lambda^{-2}$, the above integral is dominated by the 
$\vec k=0$ contribution yielding
\be
C_\sigma (t,t_w,t^*)=\frac{Y^2(t^*)}{Y(t)Y(t_w)}J\left ( \frac{t+t_w}{2} \right ) C^*
\label{04.17}
\ee
where $C^*=C(\vec k=0,t^*)$. This can be rewritten as
\be
C_\sigma (t,t_w,t^*)= t_w^{-b_\sigma}f_\sigma(x,t^*)
\label{Csigma}
\ee
where for $T_F \leq T_C$
\be
f_\sigma(x,t^*) = \Upsilon(t^*)x^{-\omega/2}(x+1)^{-d/2}
\label{fsigma}
\ee
\be
\omega = \left \{ \begin{array}{ll}
        d/2-2, \qquad $for$ \qquad T_F=T_C \\
        -d/2,  \qquad $for$ \qquad T_F<T_C
        \end{array}
        \right .
        \label{omega}
        \ee
\be
b_\sigma= \left \{ \begin{array}{ll}
        d-2, \qquad $for$ \qquad T_F=T_C \\
        0,  \qquad $for$ \qquad T_F<T_C
        \end{array}
        \right .
        \label{bsigma}
        \ee
and $\Upsilon(t^*)$ is a $t^*$ dependent constant to be determined.

With a simple change of the integration variable, the thermal fluctuations contribution~(\ref{04.18}) 
can be rewritten in the scaling form
\be
C_\psi (t,t_w,t^*) = t_w^{-b_\psi}f_\psi(x,y,t^*/t_w)
\label{Cpsi}
\ee
where
\be
f_\psi(x,y,t^*/t_w)={2T_F \over (4 \pi)^{d/2}}x^{-\omega/2}
\int_{t^*/t_w}^1 \; dz \;z^\omega(x+1-2z+y)^{-d/2}
\label{fpsi}
\ee
and
\be
b_\psi= (d-2)/2.
\label{bpsi}
\ee
The autoresponse function is obtained integrating Eq.~(\ref{03.6}) over $\vec k$ 
\be
R(t,t_w)= t_w^{-(1+a)}f_R(x,y)
\label{resp}
\ee
where
\be
a=(d-2)/2
\label{resp.1}
\ee
as $b_\psi$ and with the scaling function given by
\be
f_R(x,y)= 
(4 \pi)^{-d/2}(x-1+y)^{-(1+a)}x^{-\omega/2}.
\label{f_R}
\ee

\vspace{3mm}

\underline{{\it Quench to} $T_F=T_C$}

\vspace{3mm}

\noindent In this case $b_\sigma=2b_\psi$ and 
$C_\sigma$ becomes negligible with respect $C_\psi$, when $t_w$ is sufficiently large,
both for short and for large time separations. Hence, 
Eq.~(\ref{04.14}) can be rewritten as
\be
C(t,t_w)=C_\psi (t,t_w)
\label{Cpsi.1}
\ee
where $C_\psi (t,t_w)$ is obtained by letting $t^*/t_w \rightarrow 0$ in Eq.~(\ref{fpsi}),
since the integral is well behaved at the lower limit of integration. 
Setting $f_C(x,y)=f_\psi(x,y,0)$ and rewriting the scaling function in the form
\be
f_C(x,y)= (x-1+y)^{-b_\psi}g_C(x,y)
\label{fpsi.1}
\ee
with
\be
g_C(x,y)= {T_C \over (4\pi)^{d/2}}x^{-\omega/2}\int_{0}^{2/(x-1+y)}dz(1+z)^{-d/2}
[1-(x-1+y)z/2]^{\omega}
\label{fpsi.2}
\ee
the autocorrelation function displays the multiplicative form~(\ref{5.14}).
Furthermore, as $x$ becomes large $f_C(x,y) \sim x^{-\lambda_c/2}$ with  $\lambda_c=3d/2-2$ 
and from Eqs.~(\ref{bpsi},\ref{resp.1}) follows $a_c=b_c=b_\psi$, which is in 
agreement with Eqs.~(\ref{5.15}) and~(\ref{8.3}) since in the large $N$ model $\eta=0$ and $z_c=2$.

\vspace{3mm}

\underline{{\it Quench to} $T_F < T_C$}

\vspace{3mm}

\noindent In this case the roles of $C_\sigma (t,t_w)$ and $C_\psi (t,t_w)$ are
reversed, since now $b_\psi > b_\sigma$. In the short time regime both 
contributions are stationary with
\be
C_\sigma (\tau)=2^{-d/2}\;\Upsilon(t^*)
\label{04.26}
\ee
and
\be
C_\psi (\tau)=(M_0^2-M^2)\left [(2t^*/t_0)^{1-\frac{d}{2}}
+(\tau/t_0+1)^{1-\frac{d}{2}} \right ]
\label{04.27}
\ee
which has been obtained keeping into account that, now, the integral in Eq.~(\ref{fpsi})
develops a singularity at the lower limit of integration as $t^*/t_w \rightarrow 0$.
In order to determine $\Upsilon(t^*)$, notice that Eq.~(\ref{02.9},) together with $\xi_F ^{-2}=0$
for $T_F<T_C$, requires $C_{eq}(\vec r =0,T_F)=M^2_0$. Therefore, imposing that the
equilibrium sum rule $C_\sigma (\tau=0)+C_\psi (\tau=0)=M^2_0$ be satisfied, one gets
\be
\Upsilon(t^*)= 2^{d/2}\;\left [M^2 - (M_0^2-M^2)(2t^*/t_0)^{1-\frac{d}{2}}\right ].
\label{K}
\ee
With the above expression for $\Upsilon(t^*)$, in the large time regime one has
\be
C_\sigma (t,t_w)=\left [M^2-(M_0^2-M^2)
(2t^*/t_0)^{1-\frac{d}{2}} \right ]
\left [ \frac{4x}{(x+1)^2}\right ] ^{\frac{d}{4}}
\label{04.28}
\ee
and
\be
C_\psi (t,t_w)=(M_0^2-M^2)(2t^*/t_0)^{1-\frac{d}{2}}
\left [ \frac{4x}{(x+1)^2}\right ] ^{\frac{d}{4}}.
\label{04.29}
\ee
Taking the limit $t^*/t_0 \rightarrow \infty$ the dependence on $t^*$ is eliminated, 
yielding for short time 
\be
\left \{ \begin{array}{ll}
	C_\sigma (t,t_w)=M^2 \\
	C_\psi (t,t_w) = (M_0^2-M^2)(\tau/t_0+1)^{1-\frac{d}{2}}
          \end{array}
       \right .
\label{04.31}
\ee
and for large time  
\be
\left \{ \begin{array}{ll}
C_\sigma (t,t_w)= M^2 \left [ \frac{4x}{(x+1)^2}\right ] ^{\frac{d}{4}} \\
C_\psi (t,t_w)= 0
          \end{array}
       \right .
\label{04.32}
\ee
which implies $\lambda=d/2$.
Finally, comparing with~(\ref{6.1}) and~(\ref{5.23}), the identifications
\be
C_\sigma (t,t_w)=C_{ag}(t,t_w)
\label{04.35}
\ee
and
\be
C_\psi (t,t_w)=G_{eq}(\tau,T_F )
\label{04.34}
\ee
are obtained. 
The power law decay~(\ref{04.31}) of the stationary component $C_\psi$
is a peculiarity of the large $N$ limit, since to lowest order in $1/N$
only the Goldstone modes contribute to the thermal fluctuations~\cite{Mazenko84},
yielding critical behavior for $T_F \leq T_C$.

Following the prescription outlined in section \ref{generic-R} for the construction
of the corresponding components of the response function, from Eqs.~(\ref{8.8}) and~(\ref{04.34})
one finds 
\be
R_{eq}(\tau,T_F)= (4\pi)^{-d/2}(\tau+t_0)^{-(1+a)}
\label{Rst}
\ee
and
\begin{eqnarray}
R_{ag}(t,t_w) & = & R(t,t_w) -R_{eq}(\tau,T_F) \\ \nonumber
& = &  t_w^{-(1+a)}h_R(x,y)
\label{Rag.10}
\end{eqnarray}
with
\be
h_R(x,y) = (4\pi)^{-d/2}{x^{d/4}-1 \over (x-1+y)^{1+a}}.
\label{HR}
\ee
This completes the check that the analytical solution of the model fits into the generic pattern presented in
sections \ref{generic-C} and \ref{generic-R}.

Summarising, the additive structures of $C(t,t_w)$ and $R(t,t_w)$
have been derived from the exact solution of the model and
the pair of fields $[\sigma(\vec x,t), \psi(\vec x,t)]$, taking 
$t^* \gg t_0$, provide an explicit realization of the decomposition~(\ref{6.4}),
satisfying the requirements~(\ref{6.5}) and~(\ref{6.6}).

\subsubsection{ZFC susceptibility below $T_C$}

It is quite instructive to look in some detail at the behavior of the ZFC suceptibility
in the quench to below $T_C$, since this gives the opportunity to see a concrete 
realization of the general considerations made in section~\ref{ZFC}. 
Recalling that also $\chi(t,t_w)$ is the sum of two contributions,
by integration of~(\ref{Rst}) it is straightforward
to find the stationary component 
\be
\chi _{eq}(\tau,T_F)=(4\pi)^{-d/2}{2\over (d-2)}[t_0^{1-d/2}-(\tau+t_0)^{1-d/2}]
\label{ZFC.01}
\ee
whose long time limit gives the static susceptibility, as required by Eq.~(\ref{3.024}),
since from $t_0=\Lambda^{-2}$ and Eq.~(\ref{T_C}) follows
\be
(4\pi)^{-d/2}{2\over (d-2)}t_0^{1-d/2}=(M_0^2-M^2)/T_F = M_0^2/T_C
\label{ZFC.01y}
\ee
which coincides with the definition~(\ref{2.281}) of  $\chi _{st}(T_F)$, for $\vec r=0$.

Turning to the aging component and using Eq.~(\ref{HR}), the integral~(\ref{K.4})
entering the scaling function $h_\chi(x,y)$ is given by
\be
{\cal I}(x,y) = {1 \over (4\pi)^{d/2}}\int_{1/x}^1 dz {z^{-d/4}-1 \over (1-z+y/x)^{1+a}}.
\label{ZFC.3x}
\ee 
As $y/x$ becomes small, ${\cal I}(x,y)$ remains finite for $d < 4$, while for $d \geq 4$
there is a divergence with the leading behaviors
\be
{\cal I}(x,y) \sim \left \{ \begin{array}{ll}
        \ln(x/y), \qquad $for$ \qquad d=4 \\
        (y/x)^{-(d-4)/2},  \qquad $for$ \qquad d>4.
        \end{array}
        \right .
        \label{ZFC.3y}
        \ee
Hence, from the comparison with Eqs.~(\ref{dang},\ref{K.5},\ref{K.50}) follows $d^*=4$, $c=(d-4)/2$ and
\be
a_\chi= \left \{ \begin{array}{ll}
        a=(d-2)/2, \qquad $for$ \qquad d<d^* \\
        1,  \qquad $with log corrections for$ \qquad d=d^*\\
        1, \qquad $for$ \qquad d>d^* 
        \end{array}
        \right .
        \label{achi}
        \ee
showing that the scenario presented in section \ref{ZFC}
is verified. Namely, $y$ does to act as a dangerous irrelevant variable for $d \geq d^*$, producing
the difference between the exponents $a$ and $a_{\chi}$, and $a_{\chi}$ becomes independent of 
the dimensionality for $d \geq d^*$. As explained in section \ref{ZFC},
$d^*$ plays the role of an upper dimensionality, although 
the present context bears no relationship with critical phenomena and, therefore,
the coincidence of the values $d^*=d_U$ must be regarded as fortuitous.

\begin{figure}[h!!!!]
\begin{center}
\includegraphics*[scale = 0.5,angle = -90]{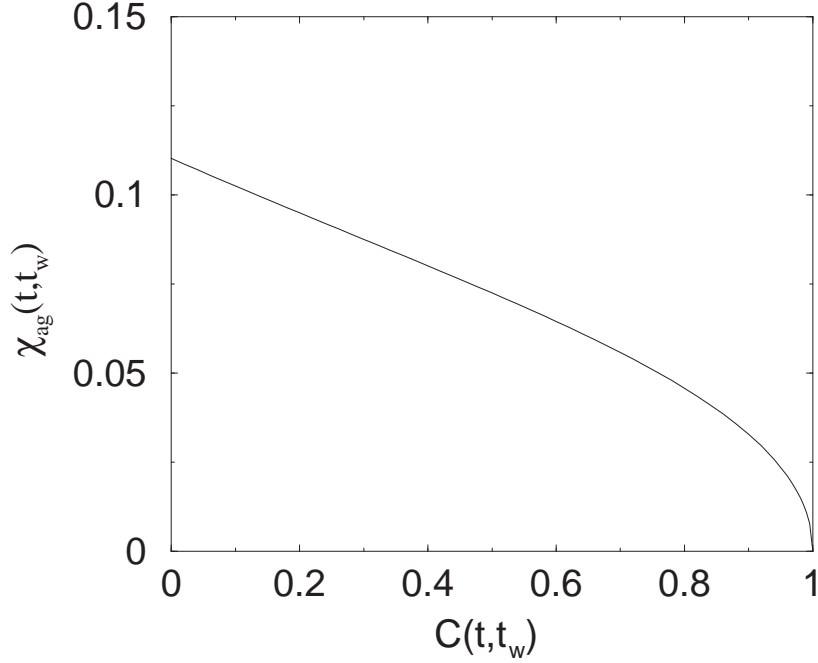}
\end{center}
\caption{\label{fig12} Parametric plot of $\chi_{ag}(t,t_w)$ in the quench of the large $N$ model 
to $(d=2,T_F=0)$.}
\end{figure}

From the above result for $a_\chi$ and from Eq.~(\ref{K.72}) follows that 
the ZFC susceptibility reaches the equilibrium value for $d>2$, while
a non vanishing value of $\chi^*$ is expected in the quench to $(d_L=2,T_F=0)$, where 
$a_\chi=0$. As $d \rightarrow 2$, from Eq.~(\ref{ZFC.01}) one has
\be
\chi _{eq}(\tau,T_F)={1 \over 4\pi}\ln(1+\tau/t_0)
\label{ZFC.6}
\ee
which diverges as $\tau \rightarrow \infty$,
as it should be since the static susceptibility~(\ref{ZFC.01y}) diverges at $d_L$, where $T_C=0$.

Switching to the aging component, the evaluation of the integral~(\ref{ZFC.3x}) for $d=2$ gives
\begin{eqnarray}
{\cal I}(x,y)& = & {1 \over 4\pi} \left [ \left ( {\sqrt{\kappa} +1 \over \sqrt{\kappa}} \right) \ln (\sqrt{\kappa} +1)
+ \left ( {\sqrt{\kappa} -1 \over \sqrt{\kappa}} \right) \ln (\sqrt{\kappa} -1) \right . \\ \nonumber
& + & \left . {1 \over \sqrt{\kappa}} \ln \left({\sqrt{\kappa} - \sqrt{1/x} \over \sqrt{\kappa} + \sqrt{1/x}} \right)
- \ln(\kappa -1/x) \right]
\label{ZFC.6x}
\end{eqnarray}
where $\kappa=1+y/x$. Hence, taking the limit $y \rightarrow 0$  one finds
\be
\chi_{ag}(x)= \hat{h}_\chi(x)= {1 \over (2\pi)} \ln \left ({2 \over 1 +x^{-1/2}} \right )
\label{ZFC.7}
\ee
and letting $x \rightarrow \infty$
\be
\chi^*= {1 \over 2\pi}\ln 2
\label{ZFC.8}
\ee
which shows that the ZFC susceptibility does not
equilibrate, although the effect is not observable since $\chi_{st}(T_F)$
diverges. As we shall see, in the $d=1$ Ising model it is not so, since the effect is observable.
The best way to visualize the formation of $\chi^*$ is through the parametric plot.
Since for $T_F=0$ the autocorrelation function is entirely given by 
$C_{ag}(t,t_w)$, eliminating $x$ between Eqs.~(\ref{04.32}) and~(\ref{ZFC.7})   
one finds
\be
\chi _{ag}(C)=\frac{1}{2\pi } \log \left \{
\frac{2}{1+\frac{M_0^2}{C}\left [ 1-\sqrt {1-\frac{C^2}{M_0^4}}\right ]}
\right \} 
\label{a2.2}
\ee
whose plot is displayed in Fig.\ref{fig12} and is qualitatively similar
to the one in Fig.\ref{fig14} for the $d=1$ Ising model. $\chi^*=0.1103$ is given by
the intercept with the vertical axis at $C=0$.

\subsection{Kinetic Ising model $d=1$}

The one dimensional kinetic Ising model with the Glauber dynamics~(\ref{3.2}) is the
other exactly soluble case~\cite{Glauber} where, as
in the large $N$ model, it is possible to derive analytically~\cite{Lippiello,GL,Sollich} the 
relaxation properties in the quench to $(d_L,T_F=0)$,
since for discrete symmetry $d_L=1$. As a matter of fact, one cannot  straightforwardly set $T_F=0$,
since, contrary to what happens for soft spins like in the GLW model, for hard spins
the linear response function $R(t,t_w)$ is not well defined for $T_F=0$.
This is due to the dependence on the temperature and on the external field entering the dynamics
through the transition rates~(\ref{3.2}) in the combination $h/T_F$, which
does not allow for a linear response regime when $T_F \rightarrow 0$. 
There is not such a problem with the Langevin equation~(\ref{3.4L}) where the temperature 
enters only through the noise, allowing to deal with a small external field also in the
zero temperature limit.
The problem can be bypassed by making the quench to a finite temperature $T_F$, 
where there are the linear regime and aging for $t_w \ll t_{eq}= \xi_F^2$, with
\be
\xi_F= -[\ln \tanh (|J|/T_F)]^{-1}.
\label{Is.01}
\ee
Hence, aging lasts longer and longer as the
temperature is lowered and $t_{eq}$ increases~\cite{Lippiello}. The behavior
in the quench to $T_F=0$, then, must be understood as the limit for $T_F \rightarrow 0$
of the off-equilibrium behavior observed in the quenches to finite $T_F$.
Alternatively, as can be seen immediately from Eq.~(\ref{Is.01}), $t_{eq}$ 
diverges by letting the coupling constant $|J| \rightarrow \infty$
while keeping $T_F$ finite and, thus, preserving the linear regime.

\begin{figure}[h!!!!]
\begin{center}
\includegraphics*[scale = 0.5,angle = -90]{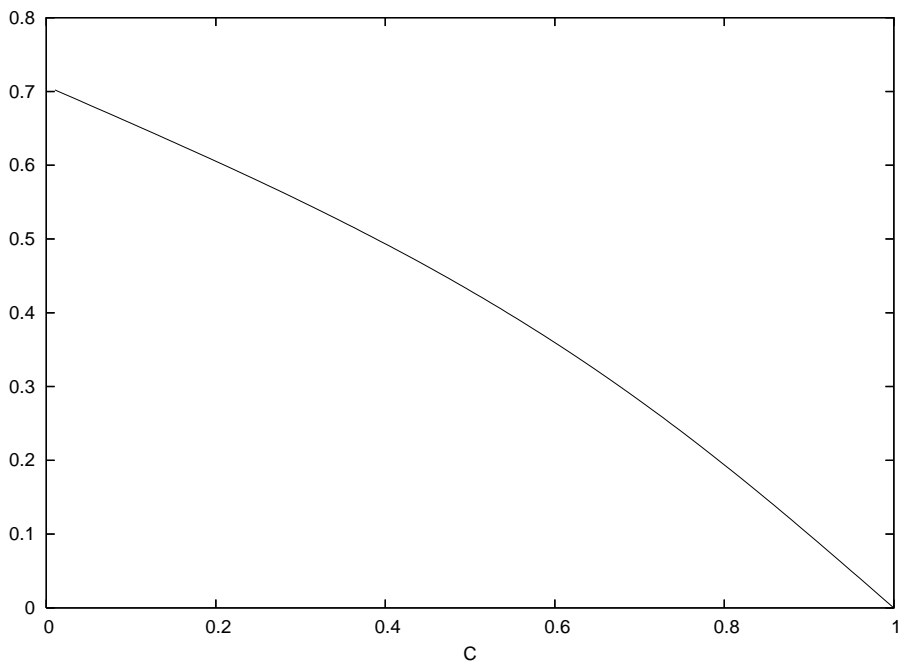}
\end{center}
\caption{\label{fig14}Plot of $T_F\chi$ against $C$ in the $d=1$ kinetic Ising model
quenched to an arbitrary value of $T_F$ with $|J|=\infty$.}
\end{figure}

\begin{figure}[h!!!!!!!]
\begin{center}
\includegraphics*[scale = 0.5,angle = -90]{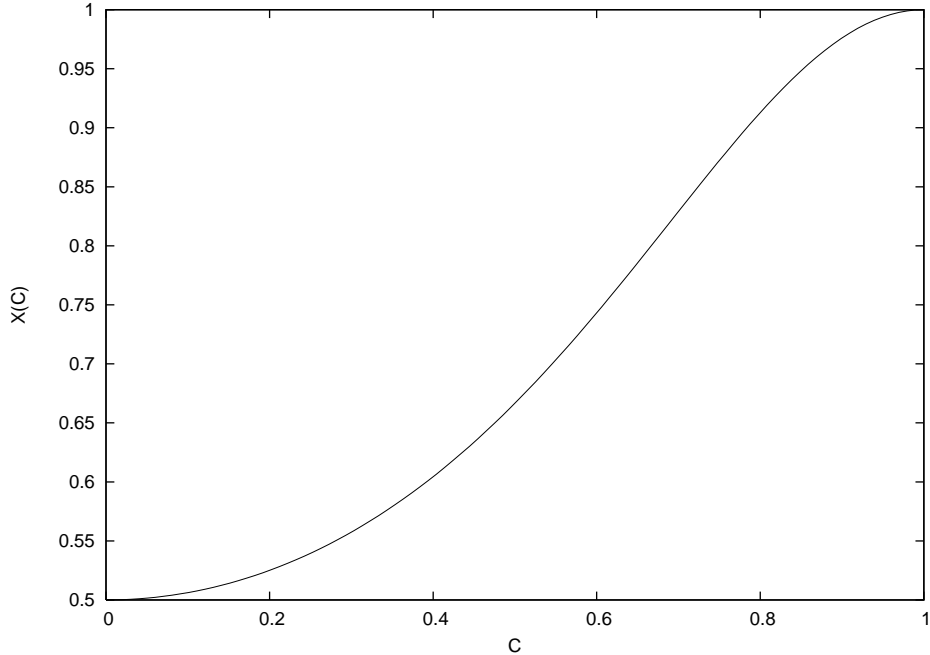}
\end{center}
\caption{\label{fig13}Parametric plot of the FDR in the $d=1$ kinetic Ising model
quenched to an arbitrary value of $T_F$ with $|J|=\infty$.}
\end{figure}

With this proviso, let us derive $R(t,t_w)$.
From the exact solution of the model~\cite{Lippiello},
the autocorrelation and autoresponse functions turn out to be related by
\be
R(t,t_w)= {1 \over 2T_F} \left [ {\partial \over \partial t_w} C(t,t_w)
- {\partial \over \partial t} C(t,t_w) \right ].
\label{Is.1}
\ee
Since for $T_F=0$, or equivalently for $|J|=\infty$, there are no thermal fluctuations
and $G_{eq}(\tau,T_F)$ vanishes,  
the autocorrelation function is entirely given by the aging component~\cite{Bray}
and reads 
\be
C(t,t_w) = {2 \over \pi} \arcsin \sqrt{\frac{2}{1+x}}.
\label{Is.3}
\ee
The same is true also for $R(t,t_w)$, since the stationary response $R_{eq}(\tau,T_F)$ vanishes at $T_F=0$
and only the aging component
gives a contribution. Inserting the above expression for $C(t,t_w)$ into Eq.~(\ref{Is.1}),
$R(t,t_w)$ is found to obey the scaling form~(\ref{I.3}) 
\be
R(t,t_w)= {1 \over \sqrt{2}\pi T_F} t_w^{-1}(x-1)^{-1/2}
\label{Is.4}
\ee
which implies $a=0$. 
Hence, as in the large $N$ model, 
it is an exact result that the exponent $a$ vanishes at $d_L$. 
The consequence, according to the general analysis
of Section~(\ref{ZFC}), is that the ZFC susceptibility does not equilibrate since the aging
contribution does not disappear in the $t_w \rightarrow \infty$ limit. 
This feature is much more conspicuous here than in the large $N$ model,
since now $\chi_{eq}(\tau,T_F)$ vanishes identically and $\chi(t,t_w)$ takes only the contribution of the aging
component, obtained by integration of Eq.~(\ref{Is.4})
\be
T_F\chi(t,t_w)={1 \over \sqrt{2} \pi} \left [ {\pi \over 2} + \arcsin \left (1-{2 \over x} \right ) \right ]
\label{Is.04}
\ee
whose $x \rightarrow \infty$ limit gives
\be
T_F\chi^*={1 \over \sqrt{2}}.
\label{Is.041}
\ee
The parametric representation, obtained eliminating $x$ with $C(t,t_w)$ in Eq.~(\ref{Is.3})
\be
T_F\widehat{\chi}(C) = {\sqrt{2} \over \pi} \arctan \left [ \sqrt{2} \cot \left (
{\pi \over 2} C \right ) \right ]
\label{Is.8}
\ee
shows (Fig.\ref{fig14}) an evident qualitative similarity with the behavior
of $\chi_{ag}(C)$ in Fig.\ref{fig12}.
Finally, the FDR is obtained by differentiating $T_F\widehat{\chi}(C)$ with respect to $C$
\be
{\cal X}(C)= \left [ 2-\sin^2\left ({\pi \over 2}C \right ) \right ]^{-1}
\label{Is.7}
\ee
offering (Fig.\ref{fig13}) an instance of the smooth and non universal behavior mentioned in
section \ref{special}.

\section{The exponent $a$}

As we have seen above, from the exact results of the large $N$ model and of the $d=1$ Ising model,
the exponent $a$ vanishes in the quenches to $(d_L,T_F=0)$. Although $a_c$, defined in
Eq.~(\ref{5.15}), vanishes as $d \rightarrow d_L$ along
the critical line, this is not an adequate explanation
because, as pointed out in section \ref{special}, the quench to $(d_L,T_F=0)$ is not a critical
quench. Hence, the vanishing of $a$ must be accounted for within the framework of the
response function in the quenches to below the critical line. Here, however, there
is an additional complication due to
a popular argument~\cite{Barrat} identifying $a$, for $T_F < T_C$, with
the exponent $n/z$ in the time dependence of the density of defects
$\rho(t) \sim t^{-n/z}$, where $n=1$ or $n=2$ for scalar or vector order parameter~\cite{Bray94}.
In order to reproduce the argument, let us rewrite Eq.~(\ref{3.020}) specializing to the
ZFC susceptibility
\be
\langle \varphi (\vec x,t) \rangle_h = \langle \varphi (\vec x,t) \rangle +
\int d\vec y \;\chi(\vec x-\vec y,t,t_w)h(\vec y) + {\cal O}(h^2)
\label{aa.1}
\ee 
and let us assume that $h(\vec x)$ is an uncorrelated random field with expectations
\be
\overline{h(\vec x)}=0
\label{aa.2}
\ee
and
\be
\overline{h(\vec x)h(\vec y)}=h_0^2 \delta(\vec x-\vec y)
\label{aa.3}
\ee
where the overbar denotes the average. Then, multiplying Eq.~(\ref{aa.1}) by $h(\vec x)$ 
and taking the average over the field one finds
\be
\chi(t,t_w) = {1 \over h_0^2} \overline{\langle \varphi (\vec x,t) \rangle_h h(\vec x)} 
\label{aa.4}
\ee 
which shows that the ZFC susceptibility is proportional to the correlation of the local 
magnetization with the external random field on the same site. Writing the
magnetization as the sum
$\langle \varphi (\vec x,t) \rangle_h
= \langle \varphi (\vec x,t) \rangle_{h,B} + \langle \varphi (\vec x,t) \rangle_{h,D}$,
where the first contribution comes from the bulk of domains, where equilibrium has been established,
and the second from the defects, the above equation takes the form
\be
\chi(t,t_w) = {1 \over h_0^2} \left [ \overline{\langle \varphi (\vec x,t) \rangle_{h,B} h(\vec x)}
+ \overline{\langle \varphi (\vec x,t) \rangle_{h,D} h(\vec x)} \right ].
\label{aa.4bis}
\ee 
Associating the two terms in the right hand side to $\chi_{eq}$ and
$\chi_{ag}$, respectively, and {\it assuming} that the defect contribution to the magnetization
is proportional to the
density of defects $\langle \varphi (\vec x,t) \rangle_{h,D} \sim \rho(t)h(\vec x)$
one finds
\be
\chi_{ag}(t,t_w) \sim \rho(t)
\label{aa.5}
\ee 
which eventually leads, recalling Eq.~(\ref{K.7}), to the identification
\be
a_\chi=n/z.
\label{aa.6}
\ee 
Since for $T_F < T_C$ the dynamical exponent $z$ is independent of
dimensionality, according to this argument also $a_\chi$ is independent of
dimensionality, implying that there is no distinction between $a_\chi$ and $a$.
Then, in the scalar case one ought to have
$a=1/2$ and in the vector case $a=1$, independently from $d$ and, therefore, also at
$d_L$.

This simple and intuitive picture is contradicted, as we have seen above, by the exact result
for the large $N$ model, which gives $a_\chi$ dependent on $d$ for $d<d^*$, and by 
the vanishing of $a$, just found, in the $d=1$ Ising model.
A point of contact with the prediction~(\ref{aa.6}) is found only
if one looks at $a_\chi$ in Eq.~(\ref{achi}) for $d > d^*$.
The question, then, is whether the large $N$ model and the $d=1$
Ising model are peculiar
cases producing exceptions to the rule~(\ref{aa.6}) or, viceversa, the 
lack of dimensionality dependence in~(\ref{aa.6}) is indicative that in
the intuitive argument some important element of the response mechanism
is missed when $d < d^*$.

In order to attempt an answer one must enrich the
phenomenology, necessarily resorting to approximate methods
and to numerical simulations. Calculations of $a$ in the scalar GLW model, with the
GAF approximation~\cite{Berthier,EPJ} and an improved version of 
it~\cite{MazenkoGAF}, give
\be
a=(d-1)/2
\label{open.3}
\ee 
which shares with the large $N$ model the linear dependence on $d$ and 
reproduces the vanishing of $a$ at $d_L=1$, as in the Ising model.
Furthermore, from the computation of the ZFC susceptibility~\cite{Berthier,EPJ} one finds for $a_\chi$
the same pattern as in Eq.~(\ref{achi}) for the large $N$ model
\be
a_\chi= \left \{ \begin{array}{ll}
        a=(d-1)/2, \qquad $for$ \qquad d<d^*\\
        1/2,  \qquad $with log corrections for$ \qquad d=d^*\\
        1/2, \qquad $for$ \qquad d> d^* 
        \end{array}
        \right .
        \label{open.4}
        \ee
except that now $d^*=2$, in place of $d^*=4$ as in the large $N$ model. The details of the analytical 
computation show
that the existence of the upper dimensionality $d^*$ occurs through the same 
mechanism as in the large $N$ model, namely with the
microscopic time $t_0$ acting as a dangerous irrelevant variable at and above $d^*$. 

Therefore, we may conclude that all the analytical results presented so far, exact and approximate, 
follow the same pattern which may be summarised by
\be
a= {n \over z}\left ({d-d_L \over d^* - d_L} \right )
\label{open.5}
\ee 
and
\be
a_\chi= \left \{ \begin{array}{ll}
        a, \qquad $for$ \qquad d<d^* \\
        n/z,  \qquad $with log corrections for$ \qquad d=d^*\\
        n/z, \qquad $for$ \qquad d>d^*
        \end{array}
        \right .
        \label{open.6}
        \ee
with $d^*$ greater than $d_L$ and dependent on the model.

\begin{figure}[h]
\begin{center}
\includegraphics*[scale = 0.5]{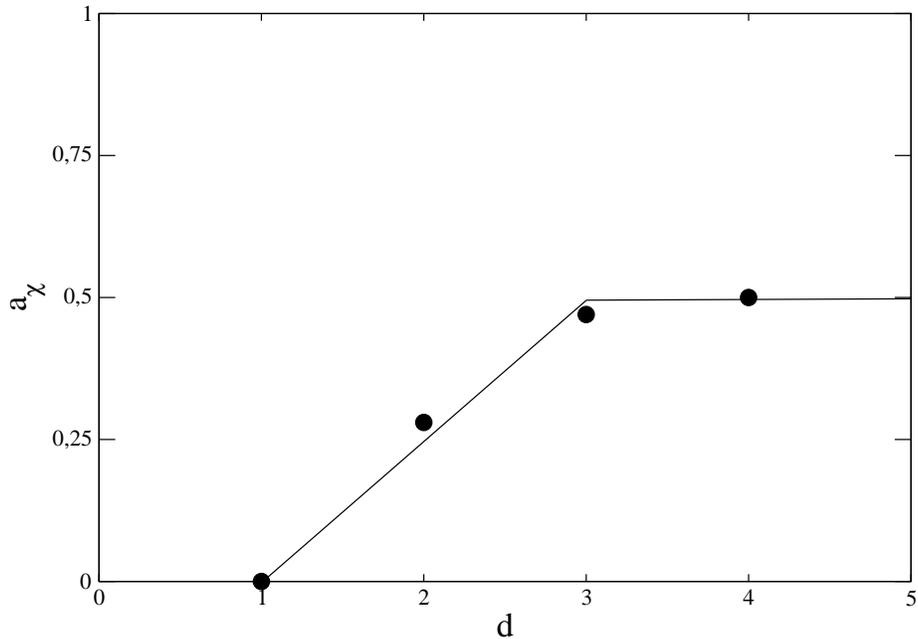}
\end{center}
\caption{\label{fig20}$a\chi$ against $d$ in the kinetic Ising model. The dot for $d=1$ is 
the exact value. The dots for $d=2,3,4$ are the results of numerical simulations. The continous
line is the plot of Eq.~(\ref{open.6}) with $n=1,z=2,d_L=1,d^*=3$. From Ref.~\cite{noiTRM}.}
\end{figure}

The next step is to see whether this pattern holds also for models accessible only
through numerical simulations. The complication with simulations is that the 
measurement of the istantaneous response
function $R(t,t_w)$ is a very demanding task from the numerical point of view.
Although significant progress has been made recently~\cite{Chat1,noialg,corrsc}, yet
a large scale survey of the behavior of $a$
upon varying $d$ in different models is unrealizable, as of now. The way out
of the numerical bottleneck is to turn to the measurement of integrated response functions,
such as the ZFC susceptibility, which are much less noisy than $R(t,t_w)$~\cite{noiTRM}.
This program has been carried out through the measurement of $a_\chi$ on systems in
different classes of universality~\cite{generic}, that is with scalar or vector order parameter,
with and without conservation of the order parameter, and at different dimensionalities.
The outcome fits into the pattern~(\ref{open.6}), provided one takes
$d^*=3$ and $d^*=4$, respectively for scalar and vector order parameter
and $z=3$ or $z=4$, for scalar conserved or vector conserved order parameter.
As an example, the values of $a_\chi$ for the non conserved Ising model with dimensionality varying from 
$d=1$ to $d=4$ are plotted in Fig.\ref{fig20}. Similar plots for other models can be found in Ref.~\cite{generic}.
For convenience the values of $d_L, d^*$ and $a_\chi$
in the two analytically treatable cases and in the Ising model are collected in the following table

\bigskip
\bigskip

\begin{tabular}{||c||c|c|c|c||}   \hline\hline
model  & $d_L$ & $d^*$ & $a_\chi=a, d \leq d^*$ & $a_\chi, d>d^*$ \\  \hline\hline
large $N$ & $2$ & $4$ & $(d-2)/2$ &  $1$  \\    \hline
GAF scalar & $1$ & $2$ & $(d-1)/2$ & $1/2$ \\    \hline
Ising numerical & $1$  & $3$ & $(d-1)/4$ & $1/2$   \\    \hline\hline 
\end{tabular}

\bigskip
\bigskip

\noindent where the expression for $a_\chi$ in the last row of the Ising case must be understood as a phenomenological formula.

Despite the good amount of evidence supporting Eqs.~(\ref{open.5}) and~(\ref{open.6}),
the issue of the exponent $a$ cannot be regarded as settled. The main reason is that a first principles derivation 
is lacking. Particularly challenging problems seem to be

\begin{enumerate}

\item what produces the deviation of $a_\chi$ from $n/z$ below $d^*$? Preliminary answers
have been put forward~\cite{HPP,generic} on the basis of the roughening of interfaces
for scalar models, but still much work remains to be done before a full understanding 
is reached.

\item in the scalar model treated with the GAF approximation, or its improved version~\cite{MazenkoGAF}, one finds $d^*=2$  
while the fit of the numerical data requires $d^*=3$.
This discrepancy reveals the strong non perturbative
nature of the exponent $a$, for which even the best analytical tools  
presently available seem to be inadequate.

\end{enumerate}

In conclusion, aging in domain growth is less trivial than commonly believed and
poses some hard problems to our understanding of phase-ordering kinetics.

\end{document}